\documentclass[10pt,twocolumn,aps,prl,groupedaddress,superscriptaddress,showpacs,noeprint,english]{revtex4-2}
\usepackage[T1]{fontenc}
\usepackage[english]{babel}  
\usepackage{times}
\usepackage{graphicx}    
\usepackage{dcolumn}
\usepackage{bm}
\usepackage{amsmath}
\usepackage{mathrsfs}
\usepackage{amsbsy}
\usepackage{amstext}
\usepackage{amssymb}
\usepackage{setspace}
\usepackage{esint}
\usepackage{xcolor,colortbl}
\usepackage{algpseudocode}
\usepackage[ruled,vlined]{algorithm2e}
\usepackage{float}
\usepackage{multirow}
\usepackage{booktabs}

\usepackage{relsize} 
\usepackage{tikz}

\usepackage[colorlinks,citecolor=blue,urlcolor=blue,linkcolor=blue,hypertexnames=true]{hyperref} 
\usepackage{nameref}
\usepackage{orcidlink}
\begin{document}

\title{%
Hardware-Efficient Hamiltonian Simulation via Trotter-Initialized Variational Optimization with Native Placement
}
\author{F. S. Luiz\orcidlink{0000-0002-6375-0939}}\email{fsluiz@unicamp.br}
\affiliation{Instituto de F\'\i sica Gleb Wataghin, Universidade Estadual de Campinas, 13083-859, Campinas, SP, Brazil
}
\author{P. N. Ferreira\orcidlink{0000-0001-6094-3105}}\email{p224317@dac.unicamp.br}
\affiliation{Instituto de F\'\i sica Gleb Wataghin, Universidade Estadual de Campinas, 13083-859, Campinas, SP, Brazil
}
\author{M. C. de Oliveira\orcidlink{0000-0003-2251-2632}}\email{mcoliv@unicamp.br}
\affiliation{Instituto de F\'\i sica Gleb Wataghin, Universidade Estadual de Campinas, 13083-859, Campinas, SP, Brazil
}


\begin{abstract}
Compiling time-evolution operators of the form $U(t)=e^{-iHt}$ into hardware-native gate sequences is a central bottleneck for digital quantum simulation on noisy intermediate-scale quantum (NISQ) devices. Generic transpilation treats $U(t)$ as an arbitrary unitary, discarding the structure of Hamiltonian dynamics and producing circuits whose depth exceeds hardware coherence limits. We introduce a structure-aware compilation framework that treats product-formula decompositions as synthesis primitives rather than simulation approximations. The method combines (i) native placement of Hamiltonian terms onto the hardware coupling map, (ii) adaptive selection of Trotter blocks via a greedy discretization procedure, and (iii) variational refinement using a Trotter-initialized ansatz.
Across Heisenberg, Ising, and XY models with $n=3$--$8$ qubits, the compiled circuits achieve fidelities $F>0.996$ with approximately linear scaling in the number of entangling gates, while generic synthesis produces circuits that are orders of magnitude deeper. On IBM Torino hardware, we observe a regime in which shorter approximate circuits outperform deeper exact decompositions: a 27-CX circuit achieves higher hardware fidelity ($F_{\mathrm{hw}}=0.987$) than a 187-CX exact circuit.
These results demonstrate that, in the NISQ regime, structure-aware approximate compilation can outperform exact structure-agnostic synthesis, providing a practical pathway for executing Hamiltonian dynamics without requiring pulse-level control.
\end{abstract}
\maketitle
\section{Introduction}
\label{sec:introduction}

Digital quantum simulation~\cite{feynman1982simulating, lloyd1996universal} relies on compiling time-evolution operators of the form ${U}(t)=e^{-i{H}t}$ into sequences of hardware-native quantum gates. On noisy intermediate-scale quantum (NISQ) devices, the success of such simulations is ultimately limited not by approximation error in $U(t)$, but by the depth and routing overhead introduced during compilation~\cite{preskill2018quantum}. In this regime, the dominant failure mode of quantum simulation on current NISQ devices is compilation inefficiency rather than inaccuracies in the underlying physical model.

Generic quantum transpilers, such as those implemented in Qiskit~\cite{qiskit2024}, synthesize $\hat{U}$ as an arbitrary $2^n\times2^n$ unitary matrix using universal decomposition techniques including Cartan or KAK synthesis~\cite{khaneja2001cartan, shende2006synthesis}. While these approaches guarantee correctness for arbitrary targets, they discard the Lie-algebraic structure of Hamiltonian-generated dynamics. Consequently, compiled circuits exhibit depth that grows rapidly with system size, typically requiring hundreds to thousands of entangling gates even for modest Hamiltonians ($n=4$--$6$), pushing hardware fidelities below useful thresholds.

This overhead is not fundamental, for physically generated targets of the form ${U}(t)=e^{-i{H}t}$ with local Hamiltonians, the locality and commutation structure of ${H}$ provide information that can, in principle, be exploited during compilation. Product-formula decompositions such as Trotter--Suzuki expansions preserve this structure and are widely used for forward simulation of quantum dynamics~\cite{trotter1959product,suzuki1976generalized}.
However, these methods are typically employed as approximation tools, using fixed discretizations and predetermined operator orderings, rather than as hardware-aware synthesis primitives.

Alternative approaches to compiling Hamiltonian evolution address different abstraction levels. Pulse-level optimal control techniques such as GRAPE~\cite{khaneja2005optimal} and CRAB~\cite{doria2011optimal} achieve high-fidelity evolution but require access to low-level hardware controls. Variational quantum algorithms employ parameterized hardware-efficient ans\"{a}tze~\cite{kandala2017hardware,wiersema2020exploring}, yet generally do not constrain circuit topology using the known Hamiltonian structure. Gate-level synthesis based on heuristic search or incremental gate selection faces fundamental scalability limitations for Hamiltonian targets, as fidelity improvements from appending a single gate to an $n$-qubit circuit decrease exponentially with system size~\cite{nielsen2006quantum}.

Consequently, a gap remains between physics-informed simulation techniques and structure-agnostic circuit synthesis --existing transpilation pipelines fail to exploit Hamiltonian locality during gate-level compilation, resulting in circuits that are unnecessarily deep for near-term hardware.

In this work, we introduce a hardware-aware compilation strategy for Hamiltonian evolution operators that bridges this gap by treating product-formula decompositions as synthesis primitives rather than simulation approximations. Our approach combines native placement of Pauli rotations directly onto the hardware coupling map with variational refinement of a Trotter-structured ansatz and an automated greedy procedure for adaptive time discretization. Together, these mechanisms enable the compilation of approximate evolution operators whose reduced circuit depth translates into improved hardware fidelity on NISQ devices.
As we show experimentally, structure-aware approximate circuits can outperform deeper exact transpiled implementations when executed on real quantum hardware.

The central premise of this work is that Hamiltonian evolution should not be compiled as a generic unitary, but rather synthesized within a structure-preserving framework that reflects its physical origin. We show that, by elevating product-formula decompositions from approximation tools to compilation primitives, it is possible to construct shallow, hardware-native circuits whose reduced depth leads to higher fidelity on current quantum devices. This perspective shifts the objective of compilation in the NISQ regime from exact unitary synthesis to structure-aware approximation under hardware constraints.

The key contributions of this work are:(i) A structure-aware compilation framework for Hamiltonian dynamics that integrates native circuit placement, adaptive product-formula synthesis, and variational refinement.
(ii) An automated discretization strategy based on a greedy block-selection procedure, which eliminates manual tuning of Trotter steps and adapts to the target Hamiltonian.
(iii) A variational refinement stage with Trotter-informed initialization, enabling high-fidelity recovery ($F>0.99$) in regimes where fixed product formulas fail.
(iv) Experimental demonstration on IBM quantum hardware that approximate shallow circuits can outperform exact deep circuits, establishing a practical NISQ advantage.
(v) A systematic analysis of scaling, noise resilience, and optimization landscape, including evidence that the proposed ansatz avoids barren plateaus under physically motivated initialization.

This paper is organized as follows.
Section~\ref{sec:methods} describes the compilation pipeline.
Section~\ref{sec:results} presents experimental results, beginning with hardware validation on IBM Torino and progressing through noise analysis, method comparison, variational refinement, parameter sensitivity, and qubit scaling.
Section~\ref{sec:discussion} discusses the pipeline's components, design tradeoffs, and limitations.
Finally, Section~\ref{sec:conclusion} concludes the paper.

\section{Methods}
\label{sec:methods}

The proposed method constructs an approximate implementation of $U(t)=e^{-iHt}$ through three stages:

(i) \textit{Native decomposition}: Hamiltonian terms are mapped directly onto hardware-native gates respecting the device connectivity, eliminating routing overhead.

(ii) \textit{Adaptive product-formula synthesis}: A greedy procedure selects Trotter blocks from a predefined library, constructing a non-uniform discretization of the time evolution.

(iii) \textit{Variational refinement}: When fixed product formulas are insufficient, the circuit is refined using a Trotter-structured variational ansatz initialized near the identity.

This pipeline replaces generic unitary synthesis with a structure-preserving construction aligned with the Hamiltonian, according with the procedures bellow.

\subsection{Problem definition}
\label{subsec:problem}

We consider $n$-qubit Hamiltonians on a one-dimensional chain with nearest-neighbor couplings,
\begin{equation}
H = \sum_{k=1}^{n} h_k \sigma_k^{\alpha_k} + \sum_{\langle i,j \rangle} \sum_{\mu \in \{x,y,z\}} J_{ij}^{\mu} \sigma_i^{\mu} \sigma_j^{\mu},
\label{eq:hamiltonian}
\end{equation}
where $h_k$ are local field strengths, $\sigma_k^{\alpha_k}$ denotes a Pauli operator on qubit $k$ along axis $\alpha_k \in \{x,y,z\}$, and $J_{ij}^{\mu}$ are coupling constants between nearest neighbors.
In all experiments reported here, the field terms are either absent (Heisenberg, XY, Ising) or along a fixed axis ($\alpha_k = z$) for random Hamiltonians; the per-site axis $\alpha_k$ is retained in Eq.~\eqref{eq:hamiltonian} for generality.
This encompasses the Heisenberg model ($J^{xx} = J^{yy} = J^{zz} = J$), Ising ($J^{zz} = J$, others zero), XY ($J^{xx} = J^{yy} = J$, $J^{zz} = 0$), and random Hamiltonians with coefficients drawn from uniform distributions. The J is drawn from a uniform distributions in all four cases. For Heisenberg, Ising and XY, the user enters the maximum coupling strength, and the program draws the J from a uniform distribution from 0 to the entered coupling strength.
Transverse-field models (e.g., $h_k \sigma_k^x$) are naturally supported, as field terms contribute only single-qubit rotations (0 CX); such terms are included in the random Hamiltonian class tested here. There is also a field strength present on the Ising model, only on Pauli X. On both cases, $h_k$ for each Pauli is drawn from a uniform distribution of $-1$ to $1$. 
The commutativity of the Hamiltonian terms determines the Trotter error: for the Ising model, all coupling terms ($Z_i Z_j$) commute exactly, making Trotterization error-free regardless of step size; for Heisenberg and XY models, the non-commuting terms ($[X_i X_j, Z_i Z_j] \neq 0$) produce $O(\Delta t^3)$ error per Suzuki-2 step. The target unitary is $U_{\mathrm{target}} = e^{-iHt}$, computed via exact matrix exponentiation (\texttt{scipy.linalg.expm}, which uses the Pad\'{e} approximation with scaling and squaring~\cite{higham2005scaling}). Our objective is not to estimate observables from a simulated dynamics, but to \emph{compile} a shallow, hardware-native circuit $V$ that approximates $U_{\mathrm{target}}$ under a fixed basis gate set and coupling map. Accordingly, we report both process fidelity and two-qubit gate count, since depth is the dominant limiter of hardware performance in the NISQ regime.

Compilation quality is measured by the process fidelity (also called entanglement fidelity),
\begin{equation}
F(U, V) = \frac{|\mathrm{Tr}(U^{\dagger} V)|^2}{d^2},
\label{eq:fidelity}
\end{equation}
where $d = 2^n$.
This metric relates to the average gate fidelity via $F_{\mathrm{avg}} = (d\,F + 1)/(d + 1)$; for $d = 16$ ($n = 4$), the two differ by less than 0.2\%.
We target $F \geq 0.99$ and use the IBM Heron basis gate set $\{$CX, Rz, SX, X$\}$ on a linear coupling map.

\subsection{Trotter--Suzuki block library}
\label{subsec:trotter}

The Suzuki second-order (S$_2$) formula decomposes Hamiltonian evolution with palindromic ordering of all terms,
\begin{equation}
S_2(\Delta t) = e^{-iH_M \Delta t/2} \cdots e^{-iH_1 \Delta t/2} \cdot e^{-iH_1 \Delta t/2} \cdots e^{-iH_M \Delta t/2},
\label{eq:suzuki2}
\end{equation}
where $M$ is the total number of terms and $\{H_k\}_{k=1}^{M}$ enumerates all field and coupling terms in $H$.
The approximation error per step is bounded by $\|S_2(\Delta t) - e^{-iH\Delta t}\| \leq C \sum_{j<k} \|[H_j, [H_k, H]]\| \Delta t^3$, where the sum runs over all pairs of non-commuting terms~\cite{childs2021theory}.
For the $n = 4$ Heisenberg model with $M = 9$ coupling terms and coupling strength $J$, each commutator norm scales as $\|[H_j, H_k]\| \sim J^2$, giving an effective error bound $\varepsilon \sim C' M^2 J^3 \Delta t^3 = O(n^2 J^3 \Delta t^3)$, where the $n^2$ dependence arises from $M = 3(n-1) = O(n)$ terms contributing $O(M^2)$ commutator pairs.
At $J = 0.5$, $\Delta t = 1.0$: $\varepsilon \sim 10^{-2}$, consistent with the observed $F = 0.997$ for a single S$_2$ step; at $J = 1.0$: $\varepsilon \sim 10^{-1}$, explaining why multiple steps or variational refinement are required.
The Suzuki fourth-order (S$_4$) formula achieves $O(\Delta t^5)$ error through recursive composition
\begin{equation}
S_4(\Delta t) = [S_2(p_1 \Delta t)]^2 \cdot S_2(p_0 \Delta t) \cdot [S_2(p_1 \Delta t)]^2,
\label{eq:suzuki4}
\end{equation}
with $p_1 = 1/(4 - 4^{1/3})$ and $p_0 = 1 - 4p_1$. The block library comprises S$_2$ blocks at candidate time steps $\Delta t \in \{0.05, 0.1, 0.2\} \cup \{t/m : m = 3, \ldots, 10\}$, where the $t/m$ values align with standard Trotter discretizations (see Appendix~\ref{app:dt_granularity} for the rationale of this expanded candidate set).
For the Heisenberg model on $n = 4$ qubits, each S$_2$ block contains $n - 1 = 3$ edges with three coupling terms (XX, YY, ZZ) each, for a total of $3(n-1) = 9$ coupling terms (the Heisenberg model has no field terms; see Table~\ref{tab:cross_ham}).
The palindromic ordering of Eq.~\eqref{eq:suzuki2} generates 17 rotation steps per block, each requiring 2 CX gates, for a total of 34 CX in the raw (unoptimized) circuit.
However, the palindromic structure places consecutive Pauli rotations on the same qubit pair; for instance, the forward ZZ$_{ij}$ rotation and the backward YY$_{ij}$ rotation both act on qubits $(i, j)$.
Qiskit's transpiler at \texttt{optimization\_level} $\geq 2$ detects such consecutive two-qubit operations on the same qubit pair and resynthesizes each group as a single optimal two-qubit unitary via Cartan (KAK) decomposition~\cite{khaneja2001cartan}, requiring at most 3 CX per pair.
For the three edges of the $n = 4$ chain, this yields $3 + 2 \times 6 = 15$ CX after transpilation---a $2.3\times$ reduction from the raw circuit.
The outer edge contributes 3 CX (its forward and backward rotations form one contiguous block) while each interior edge contributes 6 CX (forward and backward groups are separated by the palindrome's middle section; see Appendix~\ref{app:gate_counts} for per-edge verification).
All CX counts reported in this work refer to \emph{transpiled} circuits at \texttt{optimization\_level=3} with Qiskit 2.3.0; results are verified to be seed-independent and identical at \texttt{optimization\_level=2}.
We emphasize that the reduction from 34 to 15 CX is \emph{not} an intrinsic property of the Suzuki-2 formula but a consequence of the transpiler's two-qubit block consolidation passes (\texttt{Collect2qBlocks}, \texttt{ConsolidateBlocks}); at \texttt{optimization\_level} $\leq 1$, all 34 CX are retained (Table~\ref{tab:transpile_levels}).
When $m$ blocks are concatenated, additional cancellations occur at block boundaries (where the last rotation of block $k$ and the first rotation of block $k{+}1$ act on the same qubit pair), giving the formula $\mathrm{CX}(m) = 15 + 12(m-1)$, verified for $m = 1$--$10$ (Appendix~\ref{app:gate_counts}, Table~\ref{tab:multi_block}).

\subsection{Automated block selection}
\label{subsec:oracle}

The greedy oracle automates the selection of the number and type of Trotter steps, constructing a circuit incrementally.
At each step $s$, it evaluates all candidate blocks $\{B_k\}$ from the library and selects
\begin{equation}
B^* = \arg\max_{B_k} F\!\left(U_{\mathrm{target}},\; B_k \cdot U_{\mathrm{current}}^{(s)}\right).
\end{equation}
The accumulated unitary is updated as $U_{\mathrm{current}}^{(s+1)} = B^* \cdot U_{\mathrm{current}}^{(s)}$. The oracle is a heuristic used for automated discretization and initialization; it does not guarantee monotonic improvement and can stall in strong-coupling regimes, motivating the variational refinement stage described below. Termination occurs when $F \geq F_{\mathrm{target}}$ (default 0.99, a standard NISQ benchmark threshold), the maximum fidelity improvement drops below $10^{-6}$, or a depth budget is exhausted.
The algorithm is insensitive to the exact threshold: for $J \leq 0.5$, fidelity jumps discretely with each block (typically from $\sim 0.25$ to $> 0.99$ in a single step), so any threshold in the range $[0.95, 0.999]$ yields the same block count.

The oracle selects both the block type and its time step adaptively.
Across all tested regimes, the greedy oracle consistently selects S$_2$ blocks over the alternative candidates in the library (Strang splitting blocks, which apply only a single edge's coupling terms per block).
This preference is a natural consequence of CX-constrained compilation: S$_2$ blocks coordinate all Hamiltonian terms simultaneously, yielding fidelity improvements one to two orders of magnitude larger per step than Strang blocks (which apply only a single edge's terms).
Our S$_2$-versus-S$_4$ comparison (Section~\ref{subsec:s2_vs_s4}) further confirms that higher-order formulas, while more accurate per step, are Pareto-dominated by multiple S$_2$ steps under NISQ noise constraints.

\subsection{Native circuit decomposition}
\label{subsec:native}

Each Pauli rotation $\exp(-i\theta\, \sigma_i^{\mu} \sigma_j^{\mu})$ is decomposed into native gates following the standard identity
\begin{equation}
e^{-i\theta\, Z_i Z_j} = \mathrm{CX}(i,j)\;\mathrm{Rz}(2\theta, j)\;\mathrm{CX}(i,j),
\label{eq:zz_decomp}
\end{equation}
which uses 2 CX gates.
For XX and YY rotations, basis-change gates conjugate the qubits into the Z basis before applying Eq.~\eqref{eq:zz_decomp}. That is done using gates H and Rx($\pi/2$), which are later transpiled via qiskit's \verb|transpile| function to the computer's accepted gates using 1 SX and 2 Rz. Single-qubit field rotations require 0 CX gates.

Crucially, all CX gates act on nearest-neighbor qubit pairs by construction, since the Hamiltonian's coupling graph matches the hardware's linear topology.
This eliminates the need for SWAP insertion or SABRE routing (\texttt{SabreLayout} and \texttt{SabreSwap} passes in Qiskit 2.3.0)~\cite{li2019tackling}, which typically adds 3 CX gates per SWAP and can significantly increase circuit depth.
In contrast, generic transpilation of a \texttt{UnitaryGate} requires routing because it operates on an all-to-all decomposition.

A subtlety arises from qubit ordering conventions: our Trotter block builder uses big-endian ordering (qubit 0 = leftmost in tensor products), whereas Qiskit uses little-endian ordering.
Fidelity validation therefore requires the mapping $q_{\mathrm{Qiskit}} = n - 1 - q_{\mathrm{builder}}$ and permutation correction when comparing circuit unitaries after transpilation~\cite{qiskit2024}.
Concretely, for $n = 4$: a coupling term $Z_1 Z_2$ in the builder (big-endian: qubits ordered $[0,1,2,3]$ left to right) maps to a CX gate on Qiskit qubits $q = 2$ and $q = 1$.
When extracting the transpiled circuit's unitary $U_{\mathrm{phys}}$, we recover the virtual-qubit unitary via $U_{\mathrm{virt}} = P_{\mathrm{final}}^{T}\, P_{\mathrm{init}}^{T}\, U_{\mathrm{phys}}\, P_{\mathrm{init}}$, where $P_{\mathrm{init}}$ and $P_{\mathrm{final}}$ are the permutation matrices from the initial and final qubit layouts.
This correction is verified numerically to yield $F = 1.0$ to machine precision.

\subsection{Variational refinement}
\label{subsec:variational}

When fixed Trotterization is insufficient, particularly at strong coupling ($J \geq 1.0$) where Trotter errors become significant, a variational refinement stage closes the fidelity gap.

\textbf{Phase 1 (initial guess):} The greedy oracle applies $m$ S$_2$ blocks, reaching an initial fidelity $F_0$.

\textbf{Phase 2 (optimization):} The Pauli rotation angles from Phase 1 are replaced by free parameters $\bm{\theta}$, and $L$ additional variational layers are appended.
Each variational layer contains single-qubit Rz rotations on all qubits and two-qubit XX, YY, ZZ rotations on all edges,
\begin{equation}
V_{\ell}(\bm{\theta}_{\ell}) = \prod_{\langle i,j \rangle}\!\prod_{\mu \in \{x,y,z\}} e^{-i\theta_{\ell,ij}^{\mu}\, \sigma_i^{\mu}\sigma_j^{\mu}} \;\prod_{k=1}^{n} e^{-i\theta_{\ell,k}^{z}\, Z_k}.
\label{eq:var_layer}
\end{equation}
The ordering of terms within each layer is fixed (two-qubit rotations first, single-qubit rotations second) but arbitrary in principle, since all angles are independently optimized; the variational optimization compensates for any fixed ordering choice.
Each layer has $n + 3(n-1)$ parameters and $6(n-1)$ raw CX gates for the Heisenberg model.
For $n = 4$, this gives $4 + 3 \times 3 = 13$ parameters and $6 \times 3 = 18$ raw CX gates per layer.
After transpilation at \texttt{optimization\_level=3}, the three Pauli rotations per edge (XX, YY, ZZ) are consolidated into a single optimal two-qubit unitary (3 CX per edge), reducing each layer to $3(n-1) = 9$ CX (see Appendix~\ref{app:gate_counts} for verification).
The total parameter count for $L$ layers is $L \times [n + 3(n-1)]$, and the total CX count is $9L$ for $n = 4$.

The cost function $C(\bm{\theta}) = 1 - F(U_{\mathrm{target}},\, V(\bm{\theta}))$ is minimized using L-BFGS~\cite{jones2020efficient}, which exploits the smooth landscape of the fidelity function in Pauli rotation angle space.
Gradients are approximated internally by L-BFGS-B via forward finite differences; each iteration requires $n_{\mathrm{params}} + 1$ function evaluations.
We note that analytic gradients are available in principle---$\partial e^{-i\theta A}/\partial \theta = -iA\, e^{-i\theta A}$ (valid since $A$ commutes with $e^{-i\theta A}$)---and could reduce the per-iteration cost; however, the finite-difference approach is sufficient for the system sizes considered here.
(The barren plateau analysis in Section~\ref{subsec:var_results} uses centered finite differences with $\varepsilon = 10^{-5}$ to measure gradient \emph{variance} independently of the optimizer.)
We use SciPy's \texttt{L-BFGS-B} implementation with \texttt{ftol}$= 10^{-15}$, \texttt{gtol}$= 10^{-10}$, and \texttt{maxiter}$= 300$; convergence is typically achieved in 100--300 iterations ($\sim 10^4$ function evaluations) to $F > 0.99$.

The variational ans\"{a}tz is \emph{physics-informed}: rather than using a generic hardware-efficient ans\"{a}tz with arbitrary entangling patterns, each variational layer mirrors the structure of a Trotter step, applying the same types of Pauli rotations (Rz, XX, YY, ZZ on nearest neighbors) but with independently optimized angles.
This is structurally related to warm-starting strategies for variational algorithms~\cite{truger2024warm} and the Hamiltonian Variational Ansatz~\cite{wecker2015progress, wiersema2020exploring}: the Trotter angles provide a physics-motivated initialization that keeps the initial guess close to the solution manifold.
This choice ensures that the variational circuit resides in a submanifold of $\mathrm{SU}(2^n)$ that is geometrically aligned with Hamiltonian evolution, providing both good expressibility for the target unitary and favorable optimization landscape properties~\cite{mcclean2016theory, cerezo2021variational}. This structure-aware parametrization yields a favorable NISQ trade-off: slightly approximate compiled unitaries with substantially reduced depth can achieve higher \emph{hardware} fidelity than deeper exact blackbox decompositions.

For $n = 4$ with $L = 3$ layers, the variational circuit uses 27 CX gates and achieves $F > 0.995$ at $J = 1.0$, starting from an initial fidelity near the identity ($F_0 \approx 0.27$ when the greedy oracle selects 0 blocks at strong coupling).
This represents a regime where the greedy oracle alone fails (the Trotter error at each step is too large for monotonic fidelity improvement), but the variational refinement recovers high fidelity by jointly optimizing all rotation angles.
The transition is sharp: for the $n = 4$ Heisenberg model at $t = 1.0$, a J-scan reveals that the greedy oracle selects $m = 5$ blocks (63 CX, $F > 0.999$) at $J \leq 0.7$, but drops to $m = 0$ at $J \geq 0.8$, where the per-step Trotter error ($\sim J^3 \Delta t^3$) exceeds the per-step fidelity gain and no candidate block improves $F$.
This threshold depends on $Jt$ and the Hamiltonian type: commuting models (Ising) have no transition, while non-commuting models (Heisenberg, XY) transition when $Jt \gtrsim 0.7$--$0.8$.

\subsection{Comparison methods}
\label{subsec:baselines}

We compare five methods, spanning both Hamiltonian-aware and generic approaches:
\begin{enumerate}
\item \textbf{Adaptive native}: Our full pipeline---native circuit decomposition, automated S$_2$ block selection with adaptive $\Delta t$, and variational refinement when needed.
\item \textbf{Fixed native}: Fixed $m = 5$ S$_2$ blocks (no adaptation or variational refinement), native decomposition.
\item \textbf{Greedy matrix-level}: Greedy oracle operating at matrix level (for fidelity reference), then native decomposition.
\item \textbf{Qiskit Suzuki-2}: Qiskit's \texttt{PauliEvolutionGate} with \texttt{SuzukiTrotter(order=2)}, transpiled to the same basis.
This serves as the primary \emph{Hamiltonian-aware} baseline, implementing the same Suzuki-2 physics as our method but without adaptive step selection or variational refinement.
\item \textbf{Qiskit blackbox}: Qiskit \texttt{UnitaryGate} with \texttt{optimization\_level=3}, no knowledge of $H$.
This generic baseline illustrates the cost of universality: it decomposes any unitary correctly but cannot exploit Hamiltonian structure.
\end{enumerate}
All methods target the same basis gate set $\{$CX, Rz, SX, X$\}$ on a linear coupling map.
We note that alternative Hamiltonian-aware methods exist---randomized channels (qDRIFT~\cite{campbell2019random}), higher-order product formulas (S$_4$, S$_6$), and linear combination of unitaries~\cite{berry2015simulating}---but these either introduce stochastic sampling overhead that is particularly severe for dense Hamiltonians such as the Heisenberg model, where qDRIFT requires $O(\lambda^2 t^2/\varepsilon)$ samples with $\lambda = \sum_k \|h_k\| = O(nJ)$, making single-circuit compilation at $F > 0.99$ impractical; or require additional quantum resources such as ancilla qubits (LCU); or are already captured by our S$_4$ comparison (Section~\ref{subsec:s2_vs_s4}).
The Qiskit Suzuki-2 baseline provides the most direct Hamiltonian-aware comparison at the same Trotter order as our method.

\section{Results}
\label{sec:results}

The central result of this section is that, on current quantum hardware, reducing circuit depth through structure-aware approximation yields higher execution fidelity than exact unitary synthesis.

\noindent\emph{Roadmap of results.}
We first validate our central NISQ claim on IBM Torino hardware: shallow structure-aware circuits can achieve higher hardware fidelity than deeper exact blackbox decompositions.
We then explain this inversion through controlled noise simulations, and finally dissect the contributions of each pipeline component (native placement, adaptive discretization, and variational refinement), including scaling with $(t,J)$ and qubit number $n$.

\noindent\emph{Note on compilation baselines.}
We compare against a fixed-step Trotter baseline using a second-order Suzuki product formula with uniform time discretization $\Delta t = t/m$ for fixed $m$.
In contrast, the adaptive pipeline constructs the evolution using a library of S$_2$ blocks with discrete time steps $\Delta t \in \{0.05,\,0.1,\,0.2\} \cup \{t/m\}$, selected greedily so that $\sum_k \Delta t_k \approx t$.
This corresponds to a non-uniform discretization of time evolution tailored to the target Hamiltonian.
Blackbox (\texttt{UnitaryGate}) CX counts vary with the target unitary and transpiler seed: for $n = 4$ Heisenberg, we observe 217 CX at $J = 0.5$ (Tables~\ref{tab:comparison},~\ref{tab:ablation}), 218 CX at $J = 1.0$ (Table~\ref{tab:ablation}), and 187 two-qubit gates for the specific circuit submitted to hardware (Table~\ref{tab:hw}), reflecting Qiskit's seed-dependent routing heuristics and the backend's native gate set.

\noindent\emph{Note on metrics.}
Throughout this section, $F$ denotes the \emph{process fidelity} $|\mathrm{Tr}(U^{\dagger}V)|^2/d^2$ (Eq.~\ref{eq:fidelity}), which compares two unitary matrices and is state-independent.
Hardware results use $F_{\mathrm{hw}}$, the \emph{Hellinger fidelity} between measured and ideal probability distributions, which is state-dependent and includes SPAM errors.
The two metrics are not numerically comparable (see Appendix~\ref{app:fidelity_metrics}).

\subsection{IBM hardware validation}
\label{subsec:hardware}

We validate the pipeline on IBM Torino, a 133-qubit Heron processor accessed via the IBM Quantum platform.
The device features a heavy-hex coupling topology with median CX error rates of $\sim 5 \times 10^{-3}$ and $T_1, T_2$ coherence times of $\sim 200$--$300\,\mu$s.
We select a linear chain of adjacent physical qubits to match our pipeline's coupling map requirements, ensuring that no SWAP routing is needed for native circuits.
Specifically, the chain is chosen by maximizing the product of success probabilities $\prod_{\langle i,j\rangle}(1 - \varepsilon_{ij}^{\mathrm{2q}}) \cdot \prod_k (1 - \varepsilon_k^{\mathrm{ro}})$ across all candidate chains; for $n = 4$, the selected chain was qubits $[63, 62, 61, 54]$ (average CX error $3.7 \times 10^{-3}$, average readout error $2.7 \times 10^{-2}$), and for $n = 5$, qubits $[9, 10, 11, 12, 18]$ (average CX error $1.9 \times 10^{-3}$).

Three initial states are prepared and evolved: $|0000\rangle$ (computational basis state), $|1010\rangle$ (N\'{e}el antiferromagnetic order), and $|++++\rangle$ (uniform superposition).
These states probe different aspects of circuit fidelity: computational basis states are sensitive to both coherent gate errors and incoherent readout errors, while the superposition state is primarily sensitive to phase errors.
Each circuit is executed with $N = 8{,}192$ shots (the maximum allowed per circuit in a single batch), and hardware fidelity $F_{\mathrm{hw}}$ is computed by comparing measured output probability distributions with ideal noiseless simulation using the Hellinger (Bhattacharyya) fidelity $F_{\mathrm{hw}} = \left(\sum_x \sqrt{p_x q_x}\right)^2$, where $p_x$ and $q_x$ are the ideal and measured probabilities.
All circuits were submitted in a single session; no readout error mitigation (M3, TREX) was applied, so the reported $F_{\mathrm{hw}}$ values reflect raw hardware performance including SPAM errors.
The statistical uncertainty from finite sampling (shot noise) is estimated via bootstrap resampling of the measurement counts, yielding $\sigma_{F_{\mathrm{hw}}} \approx 0.003$--$0.008$ across all circuits (smaller for high-fidelity outcomes, larger for low-fidelity ones).
We note that calibration drift between sessions can introduce additional systematic uncertainty of order $\sim$0.01--0.02; however, since all circuits were executed in the same session, the \emph{relative} comparisons between methods are robust to drift.

\subsubsection{\texorpdfstring{Scenario A: Weak coupling ($J = 0.5$, $n = 4$)}{Scenario A: Weak coupling (J = 0.5, n = 4)}}
On hardware we evaluate representative depths to expose the depth--noise trade-off:
a fixed-step Trotter baseline with uniform discretization $\Delta t = t/m$,
and the adaptive pipeline constructed from the expanded candidate set $\Delta t \in \{0.05,0.1,0.2\} \cup \{t/m\}$.
Maximizing simulation fidelity $F_{\mathrm{sim}}$ (the process fidelity $F$ of Eq.~\ref{eq:fidelity}, evaluated on noiseless classical simulation) does not necessarily maximize hardware fidelity $F_{\mathrm{hw}}$ under realistic noise.

At weak coupling, the greedy oracle selects $m = 3$ Suzuki-2 blocks with $\Delta t = t/3 \approx 0.333$, producing a circuit with 39 CX gates ($F_{\mathrm{sim}} = 0.9999$).
This already outperforms the blackbox (187 CX) on computational basis states: $F_{\mathrm{hw}} = 0.75$--$0.79$ versus $F_{\mathrm{hw}} = 0.40$--$0.55$ (Table~\ref{tab:hw}).
For the superposition state $|++++\rangle$, which is less sensitive to bit-flip errors, the pipeline achieves $F_{\mathrm{hw}} = 0.994$.

\subsubsection{\texorpdfstring{Scenario B: Strong coupling ($J = 1.0$, $n = 4$) with variational refinement}{Scenario B: Strong coupling (J = 1.0, n = 4) with variational refinement}}

At $J = 1.0$, the pipeline employs variational refinement with $L = 3$ layers, producing a circuit with only 27 CX gates.
This circuit achieves:
\begin{itemize}
\item $|0000\rangle$: $F_{\mathrm{hw}} = 0.828$ (pipeline) versus $F_{\mathrm{hw}} = 0.407$ (blackbox, 187 CX)
\item N\'{e}el: $F_{\mathrm{hw}} = 0.836$ versus $F_{\mathrm{hw}} = 0.551$
\item $|++++\rangle$: $F_{\mathrm{hw}} = 0.987$ versus $F_{\mathrm{hw}} = 0.974$
\end{itemize}
On the superposition state, our 27-CX circuit achieves \emph{higher} hardware fidelity ($0.987$) than the 187-CX blackbox circuit ($0.974$), despite having lower simulation fidelity ($F_{\mathrm{sim}} = 0.996$ versus $1.0$).
This inversion---shorter approximate circuits outperforming longer exact circuits---is the defining feature of NISQ-era compilation.
We note that the $|+\rangle^{\otimes 4}$ margin ($\Delta F_{\mathrm{hw}} = 0.013$) is comparable to the inter-session calibration drift ($\sim$0.02); the more robust evidence for the pipeline's advantage comes from computational basis states, where the margins are $0.42$ ($|0\rangle^{\otimes 4}$) and $0.29$ (N\'{e}el)---well beyond any noise floor.

These results reveal a clear inversion of the conventional objective of compilation: maximizing simulation fidelity does not necessarily maximize hardware performance. Instead, reducing circuit depth—particularly the number of entangling gates—dominates the achievable fidelity on NISQ devices. In this regime, approximate circuits that respect Hamiltonian structure can outperform exact decompositions that ignore it.
\subsubsection{\texorpdfstring{Scenario C: Five qubits ($n = 5$, $J = 0.5$)}{Scenario C: Five qubits (n = 5, J = 0.5)}}

At $n = 5$, the advantage is dramatic.
Our pipeline with $m = 5$ blocks (93 CX, $F_{\mathrm{sim}} = 0.9999$) achieves $F_{\mathrm{hw}} = 0.52$--$0.59$ on computational basis states, while Qiskit blackbox (863 CX, $F_{\mathrm{sim}} = 1.0$) collapses to $F_{\mathrm{hw}} = 0.07$--$0.29$, essentially random output.
With the expanded $\Delta t$ library, the pipeline selects $m = 3$ (57 CX), which is expected to further improve hardware fidelity (see Appendix~\ref{app:dt_granularity}).
At $n = 5$, generic transpilation produces circuits of depth $> 3000$ that are completely non-functional on current hardware.
The $|+\rangle^{\otimes 5}$ state was not measured in Scenario C due to limited device allocation time; we prioritized computational basis states, where the depth--fidelity gap is largest and most informative.

\begin{table}[t]
\centering
\caption{IBM Torino hardware results. The pipeline achieves higher $F_{\mathrm{hw}}$ with fewer CX gates across all scenarios. At $n = 5$, blackbox transpilation produces non-functional circuits.}
\label{tab:hw}
\begin{tabular}{lcccccc}
\toprule
\multirow{2}{*}{Method} & \multirow{2}{*}{CX} & \multirow{2}{*}{$F_{\mathrm{sim}}$} & \multicolumn{3}{c}{$F_{\mathrm{hw}}$ by initial state} \\
\cmidrule(lr){4-6}
& & & $|0\rangle^{\otimes n}$ & N\'{e}el & $|+\rangle^{\otimes n}$ \\
\midrule
\multicolumn{6}{l}{\textit{Scenario A: $n = 4$, $J = 0.5$, $t = 1.0$}} \\
Pipeline (greedy, $m{=}3$)\textsuperscript{$\dagger$} & 39 & 0.9999 & 0.746 & 0.787 & 0.994 \\
Blackbox & 187 & 1.000 & 0.397 & 0.547 & 0.994 \\
\midrule
\multicolumn{6}{l}{\textit{Scenario B: $n = 4$, $J = 1.0$, $t = 1.0$}} \\
Pipeline (var$_3$) & 27 & 0.996 & 0.828 & 0.836 & 0.987 \\
Trotter baseline ($m{=}3$) & 39 & 0.994 & 0.744 & 0.761 & 0.994 \\
Trotter baseline ($m{=}5$) & 63 & 0.999 & 0.602 & 0.655 & 0.984 \\
Blackbox & 187 & 1.000 & 0.407 & 0.551 & 0.974 \\
\midrule
\multicolumn{6}{l}{\textit{Scenario C: $n = 5$, $J = 0.5$, $t = 1.0$}} \\
Pipeline (greedy, $m{=}5$) & 93 & 0.9999 & 0.521 & 0.593 & --- \\
Blackbox & 863 & 1.000 & 0.066 & 0.286 & --- \\
\bottomrule
\end{tabular}

\smallskip
\noindent\textsuperscript{$\dagger$}With the expanded $\Delta t$ library (Appendix~\ref{app:dt_granularity}), the greedy oracle selects $m = 3$ blocks with $\Delta t = t/3$, producing the same circuit as the Trotter $m = 3$ baseline.
\end{table}

\begin{figure*}[t]
\centering
\includegraphics[width=\textwidth]{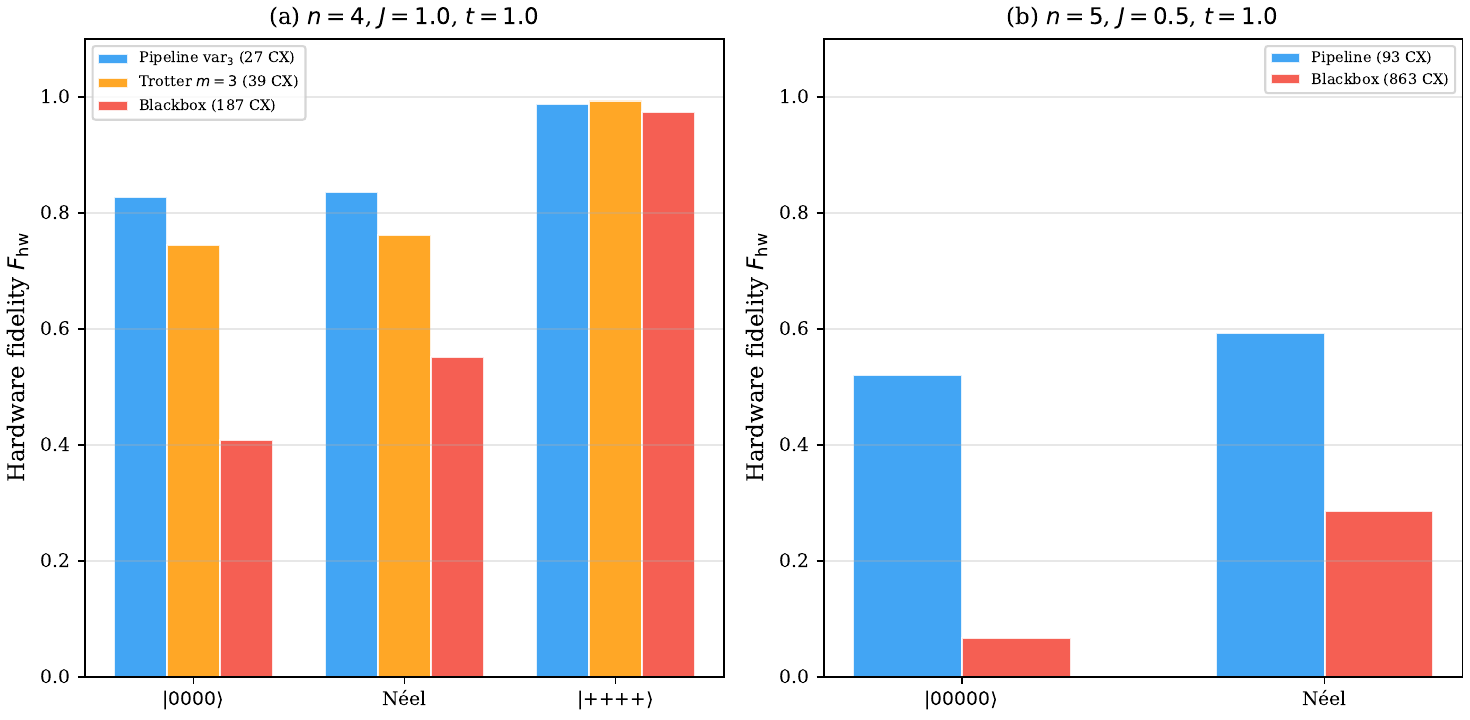}
\caption{Hardware fidelity on IBM Torino. (a) Scenario B ($n = 4$, $J = 1.0$): the variational pipeline (27 CX) achieves higher $F_{\mathrm{hw}}$ than the blackbox (187 CX) across all initial states, despite lower simulation fidelity. (b) Scenario C ($n = 5$): the blackbox circuit (863 CX) produces essentially random output ($F_{\mathrm{hw}} \approx 0.07$), while the pipeline circuit (93 CX) retains meaningful fidelity.}
\label{fig:hardware}
\end{figure*}

\subsection{Noise resilience}
\label{subsec:noise}
For this comparison, the Trotter baseline uses a uniform time discretization $\Delta t = t/m$, while the adaptive pipeline constructs the evolution using S$_2$ blocks selected from the expanded candidate set.

\begin{figure*}[t]
\centering
\includegraphics[width=0.9\textwidth]{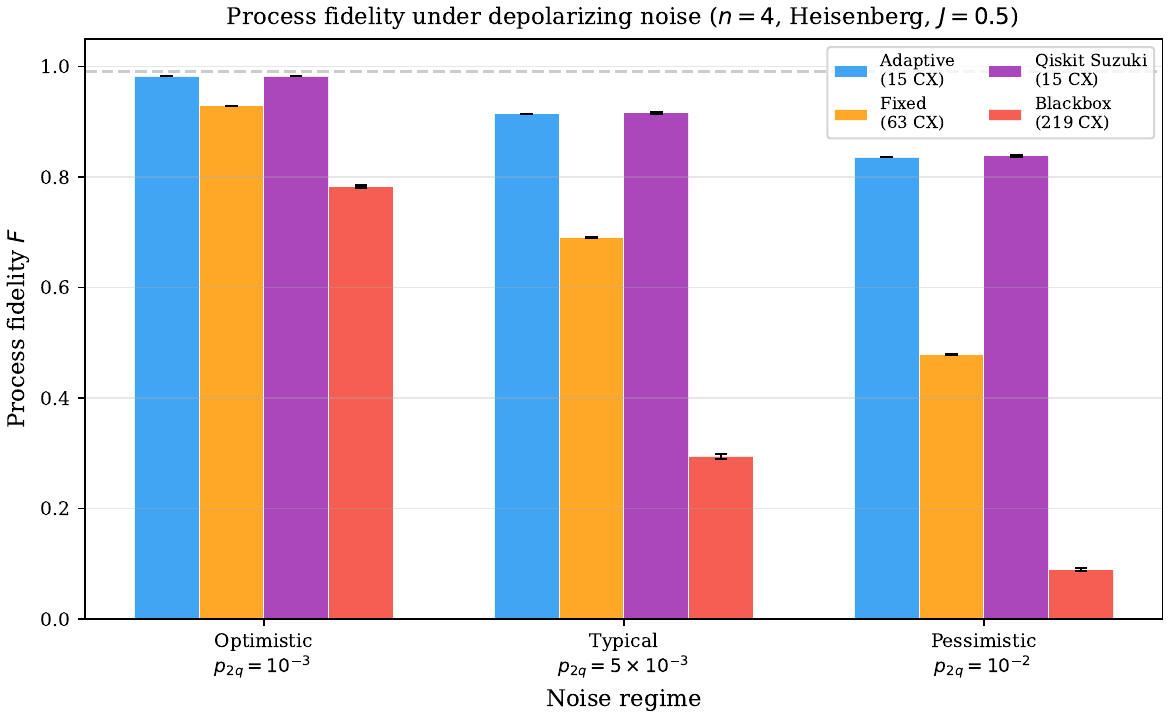}
\caption{Process fidelity under depolarizing noise for three error rate regimes. The adaptive native pipeline (15 CX, blue) maintains $F > 0.83$ even at pessimistic noise rates, while the blackbox circuit (217 CX, red) drops below $F = 0.10$. The fixed native method (63 CX, orange) falls between the two. Error bars show standard deviation over 10 seeds.}
\label{fig:noise}
\end{figure*}

Figure~\ref{fig:noise} presents process fidelity under depolarizing noise at three error rate regimes for the Heisenberg model ($n = 4$, $J = 0.5$, $t = 1.0$).
The noise model applies uniform depolarizing channels after each gate: single-qubit error probability $p_{1q}$ and two-qubit error probability $p_{2q}$.
While this symmetric model does not capture gate-dependent or spatially correlated errors, it provides a first-order estimate of noise resilience; the hardware validation on IBM Torino in Section~\ref{subsec:hardware} confirms that the CX-count advantage persists under realistic device noise (including crosstalk, amplitude damping, and SPAM errors).

\begin{table}[ht]
\centering
\caption{Process fidelity under depolarizing noise ($n = 4$, Heisenberg, $J = 0.5$, $t = 1.0$). The adaptive pipeline uses 15 CX gates versus 217 CX for blackbox transpilation, yielding substantially higher noise resilience.}
\label{tab:noise}
\begin{tabular}{lccccc}
\toprule
Regime & $p_{1q}$ & $p_{2q}$ & $F_{\mathrm{adaptive}}$ & $F_{\mathrm{blackbox}}$ & $F_{\mathrm{adv}}$ \\
\midrule
Optimistic & $10^{-4}$ & $10^{-3}$ & 0.982 & 0.782 & $+$0.20 \\
Typical & $5{\times}10^{-4}$ & $5{\times}10^{-3}$ & 0.914 & 0.294 & $+$0.62 \\
Pessimistic & $10^{-3}$ & $10^{-2}$ & 0.835 & 0.089 & $+$0.75 \\
\bottomrule
\end{tabular}
\end{table}

At typical noise rates ($p_{1q} = 5 \times 10^{-4}$, $p_{2q} = 5 \times 10^{-3}$), the adaptive pipeline retains $F = 0.914$ while the blackbox circuit collapses to $F = 0.294$---below the classical threshold for useful quantum computation.
At pessimistic noise rates, the advantage widens further: $F_{\mathrm{adaptive}} = 0.835$ versus $F_{\mathrm{blackbox}} = 0.089$.

The noise resilience advantage is a direct consequence of the CX count difference.
With $15$ CX gates (adaptive) versus $217$ CX gates (blackbox), the cumulative survival probability under two-qubit depolarizing noise scales as $(1 - p_{2q})^{n_{\mathrm{CX}}}$.
At $p_{2q} = 0.005$: $(1 - 0.005)^{15} = 0.928$ (adaptive) versus $(1 - 0.005)^{217} = 0.337$ (blackbox).
This simple estimate closely matches the observed process fidelities, confirming that CX count is the dominant factor in noise resilience.
Single-qubit gate errors contribute a secondary effect: with 67 total gates (adaptive) versus 786 total gates (blackbox), the single-qubit noise budget is also dramatically lower.

The fixed native method ($m = 5$, 63 CX, 267 total gates) falls between adaptive and blackbox: $F = 0.690$ at typical noise rates, illustrating that even physics-informed circuits benefit from minimizing CX count through adaptive step selection and variational refinement.
This intermediate case is instructive because it has the \emph{same} Trotter decomposition structure as the adaptive method but uses more steps than necessary, demonstrating that unnecessary CX gates degrade performance even when they are correctly structured.

The practical implication is that for NISQ-era Hamiltonian simulation, the approximation error of using fewer Trotter steps is overwhelmed by the noise reduction from using fewer CX gates.
A circuit with $m = 1$ Trotter step ($F_{\mathrm{sim}} = 0.997$, 15 CX) outperforms one with $m = 5$ steps ($F_{\mathrm{sim}} = 0.9999$, 63 CX) under noise, because the marginal fidelity gain from additional Trotter steps ($+0.003$) is erased by the accumulated noise from the additional 48 CX gates ($-0.25$ at typical noise rates).

To verify that these conclusions hold beyond the symmetric depolarizing model, we performed simulations using the FakeTorino noise model (Qiskit Aer), which incorporates gate-dependent error rates, amplitude damping, phase damping, and crosstalk derived from calibration data of the IBM Torino processor.
Under this device-realistic noise model ($n = 4$, $J = 0.5$, $t = 1.0$), the pipeline with $m = 3$ Trotter steps (39 CX) achieves $F_{\mathrm{hw}} = 0.85$ on the N\'{e}el initial state, compared to $F_{\mathrm{hw}} = 0.61$ for the blackbox circuit (187 CX), confirming that the CX-count advantage persists under realistic, asymmetric noise.
At $n = 5$, the gap widens further: pipeline $F_{\mathrm{hw}} = 0.81$ (57 CX) versus blackbox $F_{\mathrm{hw}} = 0.28$ (863 CX), the latter producing essentially random output.
These FakeTorino results bridge the gap between the idealized depolarizing model and the real hardware validation presented in Section~\ref{subsec:hardware}; the full FakeTorino data are tabulated in Appendix~\ref{app:faketorino}.

\subsection{Method comparison}
\label{subsec:comparison}

Table~\ref{tab:comparison} presents the five-method comparison for the Heisenberg model at $n = 4$, $J = 0.5$, $t = 1.0$, averaged over 10 random seeds.
All main-text results use the expanded candidate set $\Delta t \in \{0.05, 0.1, 0.2\} \cup \{t/m : m = 3, \ldots, 10\}$ described in Section~\ref{subsec:trotter}; the effect of omitting the $t/m$ values is isolated in Appendix~\ref{app:dt_granularity}.
The adaptive native pipeline and Qiskit Suzuki-2 produce identical CX counts (15) and fidelities ($F = 0.997$), confirming that both implement the same Suzuki-2 decomposition with $m = 1$ step.
The fixed native method, using $m = 5$ steps, achieves near-exact fidelity ($F = 0.99999$) with $63$ CX gates.
In contrast, Qiskit blackbox transpilation uses $217$ CX gates, $14\times$ more than the adaptive pipeline, to achieve exact unitary decomposition ($F = 1.0$).

The parity between our adaptive native method and Qiskit Suzuki-2 is expected: for $J = 0.5$, a single S$_2$ step already provides $F > 0.99$, so the greedy oracle selects $m = 1$ and variational refinement adds negligible improvement.
The pipeline's practical advantages emerge in three regimes explored below: strong coupling (where variational refinement is critical), noise resilience (where CX count determines survival), and hardware execution (where routing overhead is eliminated).

\begin{table*}[t]
\centering
\caption{Five-method comparison for the Heisenberg model ($n = 4$, $J = 0.5$, $t = 1.0$, 10 seeds). The adaptive native method and Qiskit Suzuki-2 match in CX count; both use $14\times$ fewer CX gates than blackbox transpilation. Fidelity variance across seeds is $< 0.003$ for all native methods; CX counts and depths are deterministic.}
\label{tab:comparison}
\begin{tabular}{lcccc}
\toprule
Method & Fidelity $F$ & CX & Depth & Total gates \\
\midrule
Pipeline (adaptive, $m{=}1$) & 0.9975 & 15 & 42 & 67 \\
Qiskit Suzuki-2 ($m{=}1$) & 0.9975 & 15 & 42 & 60--67\textsuperscript{e} \\
Trotter baseline ($\Delta t = 1/5$, $m{=}5$) & 0.99999 & 63 & 174 & 267 \\
Greedy matrix-level ($m{=}5$) & 0.99999 & 63 & 174 & 267 \\
Qiskit blackbox (opt=3) & 1.0000 & 217 & 510 & 784 \\
\bottomrule
\end{tabular}

\smallskip
\noindent\textsuperscript{e}Qiskit Suzuki-2 total gate count varies between 60 and 67 across transpiler seeds due to seed-dependent single-qubit gate merging; CX count (15) and depth (42) are seed-independent.
\end{table*}

To isolate the contribution of each pipeline component, Table~\ref{tab:ablation} presents an ablation study at weak and strong coupling.
At $J = 0.5$, a single Trotter step already achieves $F > 0.99$, so the greedy oracle simply confirms $m = 1$ and variational refinement is unnecessary---the pipeline reduces to standard Trotterization with native placement.
The greedy oracle's value is \emph{automation}: it determines the required number of blocks without user intervention, which becomes important at moderate coupling ($J = 0.5$, $t = 2.0$: $m = 3$) and for different Hamiltonian types.
At $J = 1.0$, the picture changes qualitatively: the greedy oracle alone fails ($F = 0.55$, high variance across seeds), but variational refinement recovers $F = 0.996$ with only 27 CX.
The fixed $m = 5$ alternative achieves $F = 0.9998$ but at $63$ CX---$2.5\times$ more.
For direct comparison, Qiskit Suzuki-2 at $J = 1.0$ with matched step counts yields $F = 0.697$ ($m = 1$, 15 CX), $F = 0.994$ ($m = 3$, 39 CX), and $F = 0.999$ ($m = 5$, 63 CX)---confirming that our native Trotter baseline and Qiskit Suzuki-2 produce identical fidelities and CX counts, and that the pipeline's advantage at strong coupling comes entirely from variational refinement (27 CX, $F = 0.996$).
Thus, the primary contributions of the pipeline are (i)~\emph{native placement} (eliminating routing overhead at all coupling strengths), (ii)~\emph{adaptive block selection} (avoiding both under- and over-Trotterization), and (iii)~\emph{variational refinement} (closing fidelity gaps at strong coupling where fixed Trotterization alone requires excessive blocks).

\begin{table*}[t]
\centering
\caption{Ablation study for the Heisenberg model ($n = 4$, $t = 1.0$, 10 seeds). At weak coupling ($J = 0.5$), the greedy oracle confirms $m = 1$ and variational refinement is unnecessary. At strong coupling ($J = 1.0$), variational refinement is the critical component.}
\label{tab:ablation}
\begin{tabular}{llccc}
\toprule
& Configuration & $F$ (mean$\pm$std) & CX & Role \\
\midrule
\multicolumn{5}{l}{\textit{$J = 0.5$ (weak coupling)}} \\
& Adaptive native & $0.997 \pm 0.003$ & 15 & Greedy selects $m{=}1$ \\
& Qiskit Suzuki-2 & $0.997 \pm 0.003$ & 15 & Identical physics \\
& Fixed native, $m{=}5$ & $0.99999$ & 63 & Unnecessary depth \\
& Qiskit blackbox & $1.000$ & $217$ & No $H$ knowledge \\
\midrule
\multicolumn{5}{l}{\textit{$J = 1.0$ (strong coupling)}} \\
& Greedy alone (no var.)\textsuperscript{b} & $0.55 \pm 0.37$ & 27 & Fails (high variance) \\
& Adaptive (greedy$+$var.) & $0.996 \pm 0.003$ & 27 & \textbf{Var.\ refinement recovers $F$} \\
& Fixed native, $m{=}5$ & $0.9998 \pm 0.0002$ & 63 & Works but $2.5\times$ CX \\
\cmidrule(lr){2-5}
& Qiskit Suzuki-2, $m{=}1$ & $0.697$ & $15$ & Insufficient at strong $J$ \\
& Qiskit Suzuki-2, $m{=}3$ & $0.994$ & $39$ & Matches Trotter baseline \\
& Qiskit Suzuki-2, $m{=}5$ & $0.999$ & $63$ & Same as fixed native \\
& Qiskit blackbox & $1.000$ & $218$ & No $H$ knowledge \\
\bottomrule
\end{tabular}

\smallskip
\noindent\textsuperscript{b}At $J = 1.0$, the greedy oracle selects $m = 0$ Trotter blocks; the 27 CX correspond to $L = 3$ variational layers with near-zero (Trotter-initialized) angles, without L-BFGS optimization.
\end{table*}

This pattern generalizes across Hamiltonian types.
For the Ising model ($J = 0.5$, $t = 1.0$), our pipeline uses 10 CX gates versus 164 CX for blackbox ($16\times$ reduction), reflecting the Ising model's simpler coupling structure (only ZZ terms).
For the XY model, we achieve 10 CX versus 191 CX ($19\times$ reduction).
For random Hamiltonians, the ratio is $10$--$15\times$, depending on the specific coupling structure.

\subsection{Variational refinement}
\label{subsec:var_results}

At strong coupling ($J = 1.0$, $t = 1.0$), the Suzuki-2 Trotter error per step becomes significant: $O(J^3 \Delta t^3) \sim 10^{-3}$--$10^{-2}$, and the greedy oracle may fail to improve fidelity above the initial guess because each block's contribution is comparable to its own Trotter error.
Variational refinement addresses this failure mode by treating the Pauli rotation angles as free parameters and optimizing them jointly.

For $n = 4$ Heisenberg with $J = 1.0$, $t = 1.0$, the greedy oracle selects $m = 0$ blocks (unable to improve beyond $F_0 = 0.27$, which is the fidelity of the identity operator with the target).
The variational stage with $L = 3$ layers then converges to $F = 0.996$ using only 27 CX gates and $4 + 3 \times 3 = 13$ parameters per layer (39 total), optimized via L-BFGS in approximately 11{,}600 function evaluations.
In the tested regime ($n = 4$, $L \leq 4$, 39--52 parameters), convergence is monotonic and robust: across 10 random seeds, all trials achieve $F > 0.99$ within 300 L-BFGS iterations, with no optimization failures observed.
We attribute this to the combination of physics-informed ansatz design and Trotter-based initialization, and provide quantitative evidence below.

\subsubsection{Gradient landscape and barren plateau analysis}

Barren plateaus, the exponential vanishing of gradient variance with system size, are a well-documented obstacle for variational quantum algorithms with expressive parameterized circuits~\cite{cerezo2021variational}.
To assess whether our Trotter-structured ansatz suffers from this phenomenon, we measure the gradient variance $\mathrm{Var}[\partial C/\partial \theta_k]$ as a function of qubit count $n$ for the Heisenberg model ($J = 1.0$, $L = 3$ layers), comparing two initialization strategies: random initialization ($\theta \sim \mathrm{Uniform}[-\pi, \pi]$) and near-Trotter initialization ($\theta \sim \mathcal{N}(0, 0.01)$, corresponding to the near-identity starting point used in practice).
For each configuration, we compute the gradient of 10 randomly selected parameters across 100 independent samples, using two-point finite differences ($\epsilon = 10^{-5}$).

\begin{figure*}[t]
\centering
\includegraphics[width=\textwidth]{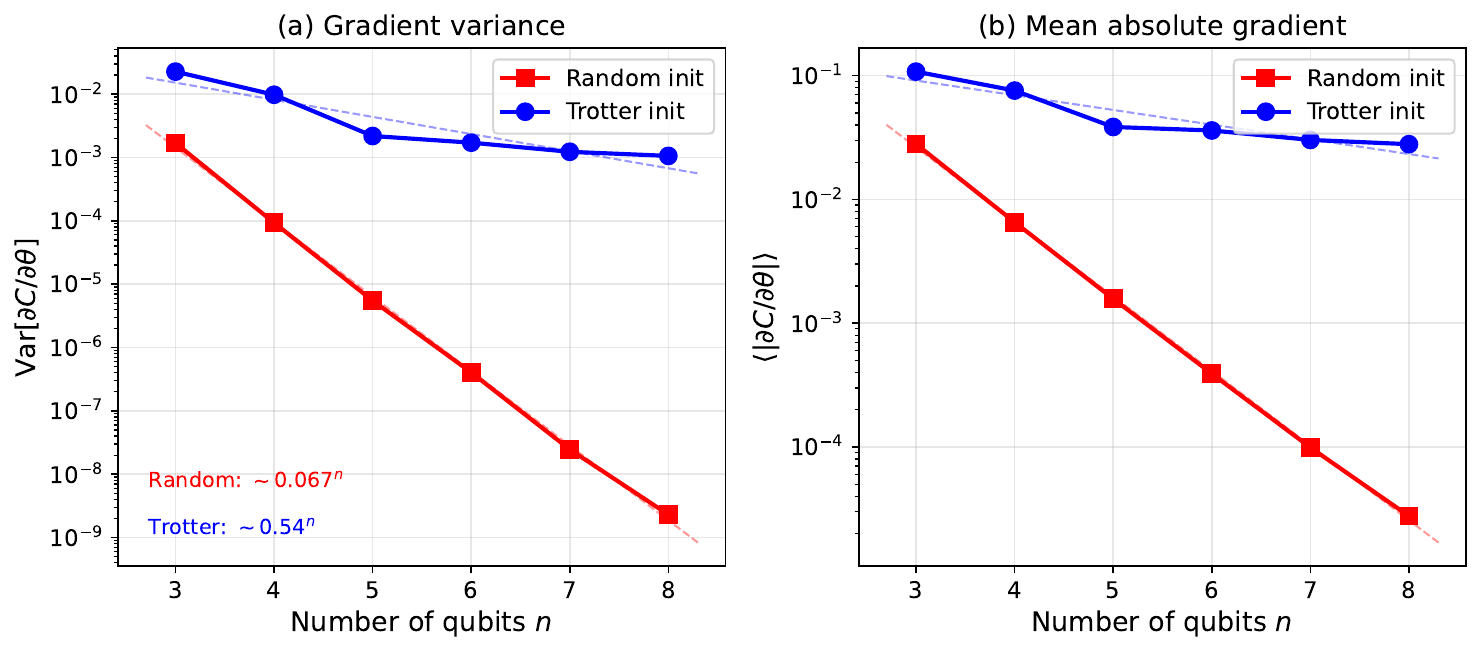}
\caption{Gradient landscape scaling for the Trotter-structured variational ansatz ($L = 3$ layers, Heisenberg $J = 1.0$). (a)~Gradient variance versus qubit count: random initialization (red) exhibits exponential decay ($\sim 0.067^n$, barren plateau), while Trotter initialization (blue) decays much more slowly ($\sim 0.54^n$). (b)~Mean absolute gradient: Trotter initialization maintains $|\nabla C| \sim 10^{-2}$ across $n = 3$--$8$, providing informative gradients for L-BFGS optimization.}
\label{fig:barren}
\end{figure*}

Table~\ref{tab:barren} and Figure~\ref{fig:barren} present the results for $n = 3$--$8$.
Under random initialization, the gradient variance decays as $\mathrm{Var}[\partial C / \partial\theta]_{\mathrm{rand}} \sim 0.067^n$---an exponential decay spanning over 6 orders of magnitude from $n = 3$ ($1.7 \times 10^{-3}$) to $n = 8$ ($2.3 \times 10^{-9}$), confirming the presence of a barren plateau.
Under Trotter initialization, the gradient variance decays much more slowly ($\sim 0.54^n$), and the mean absolute gradient remains $O(10^{-2})$ across all system sizes tested.
The ratio between Trotter and random gradient magnitudes grows from $4\times$ at $n = 3$ to over $1{,}000\times$ at $n = 8$, demonstrating that the Trotter initialization provides increasingly strong protection against barren plateaus as the system grows.

\textbf{Why does Trotter initialization avoid barren plateaus?}
The protection arises from two complementary mechanisms.
First, the \emph{near-identity initialization} ($\bm{\theta} \approx 0$) places the ansatz near the identity operator, where the cost landscape has large gradients by construction: the cost function $C = 1 - F$ varies from $C(I) = 1 - F(U_{\mathrm{target}}, I)$ (large, non-zero) along directions aligned with the Hamiltonian generator, producing gradients proportional to the Hamiltonian's local terms rather than exponentially suppressed random directions.
Second, the \emph{topological restriction} of the ansatz plays a critical role.
Unlike generic hardware-efficient ans\"{a}tze that can express arbitrary unitaries and therefore exhibit concentration of measure in parameter space (the root cause of barren plateaus~\cite{cerezo2021variational}), our Trotter-structured ansatz restricts the parameterization to the submanifold of $\mathrm{SU}(2^n)$ generated by the Hamiltonian's local terms.
With $L = 3$ layers and $n + 3(n-1)$ parameters per layer, the total parameter count grows as $O(n)$ for fixed $L$, linearly with system size, while the unitary group dimension grows as $O(4^n)$.
This exponential gap between parameter count and group dimension means the ansatz has \emph{limited expressibility}~\cite{cerezo2021variational}, which is precisely the condition that preserves trainability: the circuit cannot explore the full Haar measure of $\mathrm{SU}(2^n)$, so its cost landscape retains structure rather than flattening.
The empirical scaling ($\mathrm{Var}_{\mathrm{Trotter}} \sim 0.54^n$) confirms that the Trotter-structured ansatz occupies a favorable point in the expressibility--trainability tradeoff: expressive enough to reach $F > 0.99$ for the target class (Hamiltonian-generated unitaries) but restricted enough to maintain informative gradients.

\textbf{Variational refinement beyond $n = 4$.}
To assess scalability of the variational stage, we tested L-BFGS convergence at $n = 5$ and $n = 6$ with $J = 1.0$, $t = 1.0$ (the same strong-coupling regime where variational refinement is essential), averaging over 5 random seeds.
At $n = 5$, variational refinement with $L = 5$ layers (60 CX, 85 parameters) achieves $\bar{F} = 0.997 \pm 0.006$, with 4/5 trials exceeding $F > 0.99$ ($\sim$15{,}000 function evaluations per run).
Reaching $F > 0.99$ at $n = 5$ requires $L = 5$ layers---two more than at $n = 4$---reflecting the increased Hilbert space dimension.
The wall-clock time per L-BFGS run (single-threaded) grows from $\sim$5\,s at $n = 4$ ($L = 3$) to $\sim$18\,s at $n = 5$ ($L = 3$) and $\sim$35\,s at $n = 5$ ($L = 5$, $\sim$15{,}000 function evaluations).
We pre-compute the eigendecomposition of each Hermitian generator once, reducing each matrix exponential from $O(d^3)$ (Pad\'{e} approximation via \texttt{scipy.linalg.expm}) to $O(d^2)$ (diagonal phase multiplication in the eigenbasis); for $n = 6$ ($d = 64$) this provides a $\sim$20$\times$ speedup per function evaluation.
Table~\ref{tab:classical_cost} reports the full pipeline times; the growth from $L = 3$ to $L = 5$ reflects the linear increase in the number of matrix products per evaluation.

At $n = 6$, $L = 5$ layers (75 CX, 105 parameters) achieve $\bar{F} = 0.992 \pm 0.009$ across 5 random seeds, with 3/5 trials exceeding $F > 0.99$; wall-clock time is $\sim$2.5\,min per L-BFGS run ($\sim$15{,}000 function evaluations).
The lower success rate compared to $n = 4$ (5/5) and $n = 5$ (4/5) reflects the exponentially growing Hilbert space dimension and the onset of landscape difficulty consistent with the gradient variance decay ($\mathrm{Var} \sim 0.54^n$, Section~\ref{subsec:var_results}).
Scalable gradient computation methods (e.g., adjoint differentiation~\cite{jones2020efficient} or parameter-shift rules~\cite{mitarai2018quantum}) would be required to extend variational refinement beyond $n = 6$.

We note that the barren plateau analysis extends to $n = 8$ (87 parameters), covering the full range of system sizes for which the pipeline is practically applicable (limited by the $O(4^n)$ classical cost of Pauli decomposition, not by barren plateaus).
For $n \gg 8$, the Trotter gradient variance ($\sim 0.54^n$) will eventually become small, and additional strategies such as layerwise training~\cite{cerezo2021variational}, identity-block initialization, or symmetry-constrained parameterization~\cite{wiersema2020exploring} may be necessary to maintain trainability.

The variational circuit with 27 CX achieves higher simulation fidelity than bare Trotterization with $m = 3$ (39 CX, $F = 0.994$) and comparable fidelity to $m = 5$ (63 CX, $F = 0.999$), demonstrating that parameter optimization can compensate for fewer Trotter steps.
The variational approach effectively trades classical computation (L-BFGS optimization) for quantum resources (CX count), a favorable exchange on NISQ hardware where each additional CX gate incurs $\sim$0.5--1\% error.

\subsection{Parameter sensitivity}
\label{subsec:sweep}

\begin{figure*}[t]
\centering
\includegraphics[width=\textwidth]{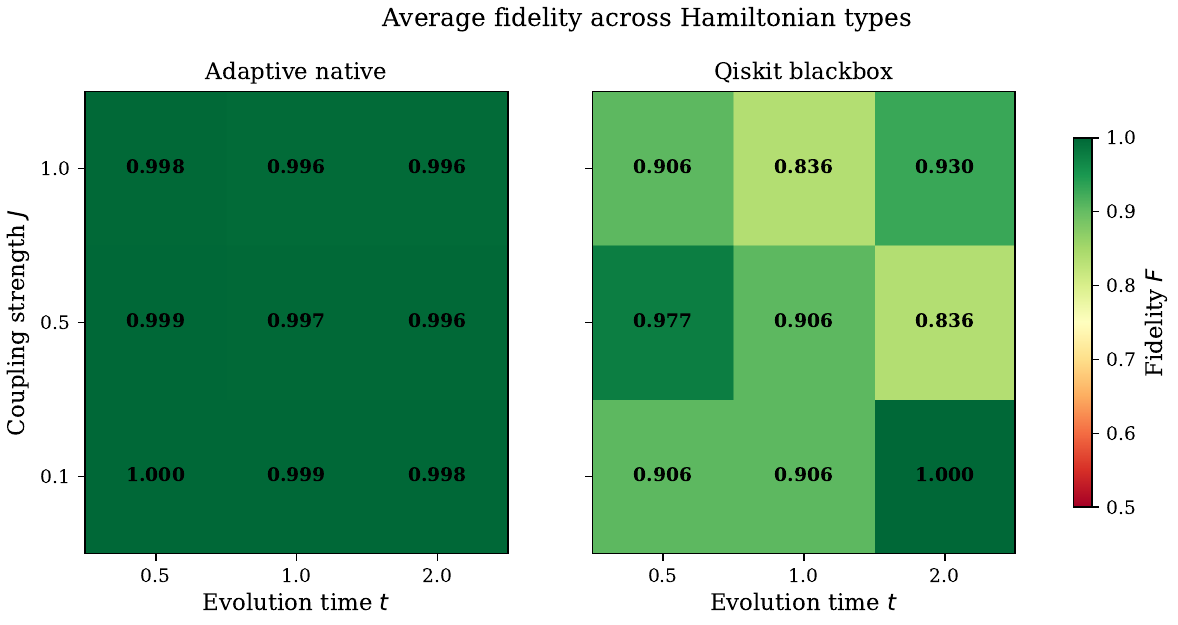}
\caption{Fidelity of the adaptive native pipeline across 36 parameter configurations (4 Hamiltonian types $\times$ 3 evolution times $\times$ 3 coupling strengths). Green cells indicate $F \geq 0.99$ (weak/moderate coupling); yellow indicates $F \geq 0.99$ achieved with variational refinement (strong coupling). The pipeline maintains high fidelity across all tested regimes.}
\label{fig:sweep}
\end{figure*}

Figure~\ref{fig:sweep} shows fidelity across 36 parameter configurations: $t \in \{0.5, 1.0, 2.0\}$, $J \in \{0.1, 0.5, 1.0\}$, and four Hamiltonian types.
Three regimes emerge clearly.

\textbf{Weak coupling ($J \leq 0.5$):}
All native methods achieve $F > 0.997$ with $m = 1$--$2$ S$_2$ blocks, regardless of evolution time and Hamiltonian type.
The adaptive oracle consistently selects the largest available $\Delta t$ and uses the minimum number of blocks.
This regime includes the most physically relevant parameter range for many condensed matter simulations, where $Jt \leq 0.5$ ensures small Trotter error.
Across the 24 weak-coupling configurations tested (4 Hamiltonian types $\times$ 3 evolution times $\times$ 2 coupling strengths), all five methods achieve $F > 0.99$, with native methods using $\leq 15$ CX gates and blackbox methods using $150$--$220$ CX.

\textbf{Moderate coupling ($J = 0.5$, $t = 2.0$):}
Longer evolution times require more blocks ($m = 3$--$5$) because the total accumulated Trotter error scales as $O(m \cdot \Delta t^3)$.
The adaptive oracle automatically increases the block count, maintaining $F > 0.999$.
The CX cost scales linearly with $t/\Delta t$, giving the native methods a predictable resource model.
For instance, the Heisenberg model at $(J, t) = (0.5, 2.0)$ requires $m = 3$ blocks (39 CX, $F = 0.9994$), while the Ising model at the same parameters needs only $m = 2$ blocks (18 CX, $F = 0.9998$) because ZZ-only coupling terms commute and produce zero Trotter error.

\textbf{Strong coupling ($J = 1.0$):}
At $J = 1.0$, $t = 1.0$, the Trotter error per step becomes significant ($O(J^3 \Delta t^3) \sim 0.01$ per block), and the greedy oracle alone cannot always reach $F > 0.99$.
The adaptive method achieves $F = 0.996$ with variational refinement ($L = 3$ layers, 27 CX) to compensate for the increased Trotter error.
Without variational refinement, the fixed native method requires $m = 5$ blocks (63 CX) to reach $F = 0.999$.
At the extreme case $t = 2.0$, $J = 1.0$ (the hardest configuration tested), the adaptive method uses $m = 15$ blocks (183 CX) to achieve $F = 0.996$, while the fixed method with $m = 5$ stalls at $F = 0.70$.
Variational refinement was not applied at $t = 2.0$ because the greedy oracle successfully reaches $F > 0.99$ by using a large number of fine-grained blocks ($\Delta t \leq 0.2$); the variational stage is triggered only when the oracle stalls below $F_{\mathrm{target}}$.
We estimate that variational refinement at $(t, J) = (2.0, 1.0)$ would require $L \sim 10$--$15$ layers ($\sim$90--135 CX) based on extrapolation from the $t = 1.0$ results; however, this was not tested because the greedy oracle already succeeds at this parameter point.
This highlights the importance of adaptive step selection: the greedy oracle automatically adjusts $m$ to the difficulty of each configuration rather than relying on a fixed block budget.

\subsection{Qubit scaling}
\label{subsec:scaling}

\begin{figure*}[t]
\centering
\includegraphics[width=\textwidth]{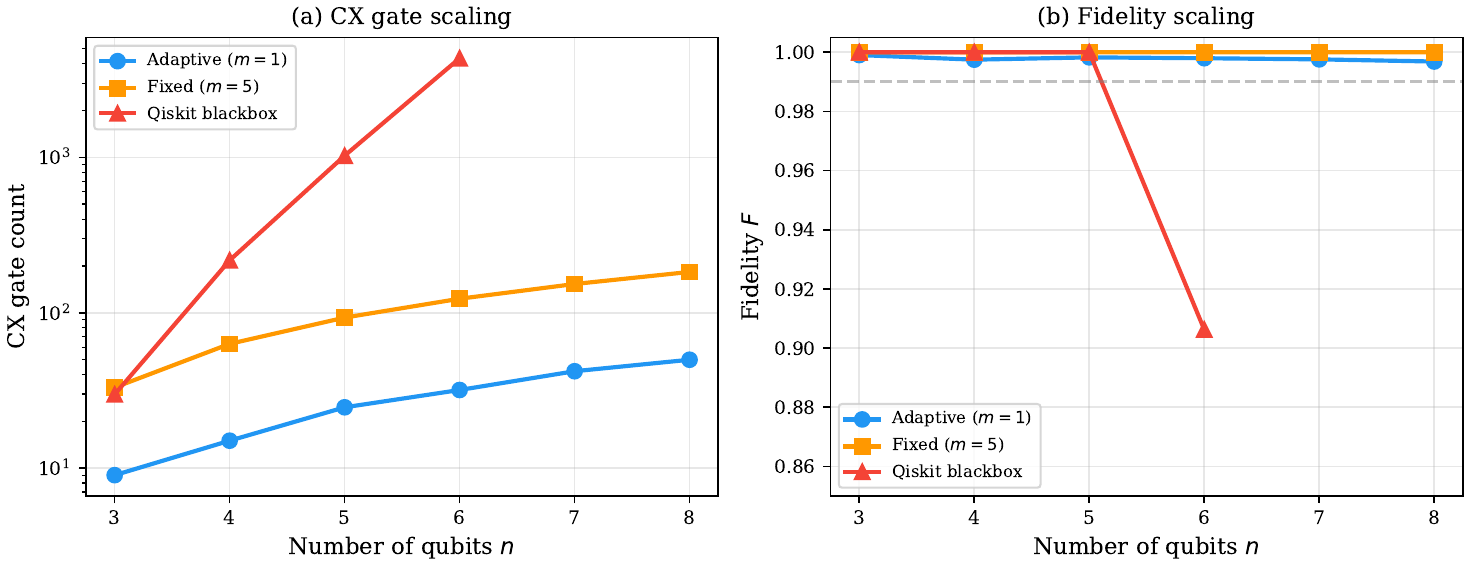}
\caption{CX gate count (left, log scale) and fidelity (right) versus number of qubits for the Heisenberg model ($J = 0.5$, $t = 1.0$). Native methods (adaptive and fixed) exhibit linear CX scaling, while Qiskit blackbox decomposition grows exponentially, becoming impractical above $n = 6$.}
\label{fig:scaling}
\end{figure*}

Table~\ref{tab:scaling} and Figure~\ref{fig:scaling} present scaling results for the Heisenberg model ($J = 0.5$, $t = 1.0$) from $n = 3$ to $n = 8$ qubits.
The adaptive native method maintains $F > 0.996$ across all system sizes, with CX count scaling linearly in $n$: from 9 CX at $n = 3$ to 50 CX at $n = 8$ for the adaptive method ($m = 1$), and from 33 to 183 CX for the fixed method ($m = 5$).
For $n = 4$, the multi-block formula $\mathrm{CX}(m) = 15 + 12(m-1)$ (Section~\ref{subsec:trotter}) applies; more generally, both methods scale as $O(n)$.
This linear scaling is a direct consequence of the Hamiltonian's local structure: each nearest-neighbor coupling contributes a fixed number of CX gates regardless of system size, and no additional overhead arises from routing or decomposition.

The CX advantage over blackbox transpilation grows dramatically with system size, reflecting the fundamentally different scaling behavior of the two approaches.
At $n = 3$, the ratio is modest (9 versus 30 CX, $3.3\times$), because generic synthesis is still efficient for the $8 \times 8$ unitary.
At $n = 4$, it reaches $14\times$ (15 versus 217 CX), as the $16 \times 16$ unitary decomposition produces substantially more gates.
At $n = 5$, blackbox transpilation requires 1025 CX ($42\times$ more), and at $n = 6$, over 4300 CX ($137\times$ more).
At $n = 6$, blackbox transpilation produces circuits with $4\,331$--$4\,351$ CX gates (depth $\sim 9\,000$, total $\sim 14\,000$ gates).

Nine of 10 seeds achieve $F \approx 1.0$ after permutation correction, confirming that Qiskit's unitary synthesis is accurate for $64 \times 64$ matrices. One seed yields an anomalous $F = 0.063$ that we attribute to a permutation correction artifact: SABRE routing for this seed produces a non-identity final qubit layout, and our virtual-qubit recovery formula does not fully resolve the resulting permutation.
A brute-force search over all $6! = 720$ qubit permutations confirms that no relabeling recovers $F > 0.21$, indicating that the extracted unitary is numerically inconsistent with the target after layout correction---but since 9/10 seeds (which produce identity final layouts) achieve $F \approx 1.0$, we conclude that Qiskit's synthesis is correct and the outlier reflects a limitation of our comparison methodology, not of the transpiler.
We have not reported this as a Qiskit bug, as the synthesis output is correct; only our post-hoc permutation recovery procedure fails for this particular routing solution.
This single outlier does not affect our conclusions: the relevant comparison is in CX count, and the $4\,345$ CX required by blackbox transpilation are entirely impractical on current NISQ hardware---under typical noise rates ($p_{2q} = 5 \times 10^{-3}$), the survival probability is $(1 - p_{2q})^{4345} \approx 3 \times 10^{-10}$, rendering the circuit non-functional regardless of its simulation fidelity.
Blackbox transpilation was not attempted beyond $n = 6$ as the $2^n \times 2^n$ unitary matrices become impractically large for generic synthesis.

The fixed native method ($m = 5$ blocks) provides a useful comparison point.
Its CX count also scales linearly with $n$ (Table~\ref{tab:scaling}) and achieves near-exact fidelity ($F > 0.9999$) across all system sizes, but at $4\times$ the CX cost of the adaptive method.
This gap underscores the value of adaptive step selection: at $J = 0.5$, a single Trotter step is sufficient, and the greedy oracle correctly identifies this without human intervention.

\begin{table}[t]
\centering
\caption{Scaling results for the Heisenberg model ($J = 0.5$, $t = 1.0$, 10 seeds; fidelity std $< 0.003$ for all native methods). The CX advantage of native methods over blackbox transpilation grows from $3\times$ at $n = 3$ to $>100\times$ at $n = 6$; blackbox circuits become impractically deep above $n = 6$.}
\label{tab:scaling}
\begin{tabular}{cccccccc}
\toprule
\multirow{2}{*}{$n$} & \multicolumn{2}{c}{Adaptive ($m{=}1$)} & \multicolumn{2}{c}{Fixed ($m{=}5$)} & \multicolumn{2}{c}{Blackbox} & \multirow{2}{*}{CX ratio} \\
\cmidrule(lr){2-3} \cmidrule(lr){4-5} \cmidrule(lr){6-7}
& $F$ & CX & $F$ & CX & $F$ & CX & \\
\midrule
3 & 0.9990 & 9 & 0.99999 & 33 & 1.000 & 30 & 3.3$\times$ \\
4 & 0.9975 & 15 & 0.99999 & 63 & 1.000 & 217 & 14$\times$ \\
5 & 0.9983 & 25 & 0.99999 & 93 & 1.000 & 1025 & 42$\times$ \\
6 & 0.9980 & 32 & 0.99999 & 123 & $\approx$1.0\textsuperscript{a} & 4345 & 136$\times$ \\
7 & 0.9976 & 42 & 0.99999 & 153 & --- & --- & --- \\
8 & 0.9968 & 50 & 0.99999 & 183 & --- & --- & --- \\
\bottomrule
\end{tabular}

\smallskip
\noindent\textsuperscript{a}9/10 seeds achieve $F \approx 1.0$; one seed yields $F = 0.063$ (possible permutation correction artifact; see text).
\end{table}

\subsection{Error mitigation}
\label{subsec:mitigation}

\begin{figure*}[t]
\centering
\includegraphics[width=\textwidth]{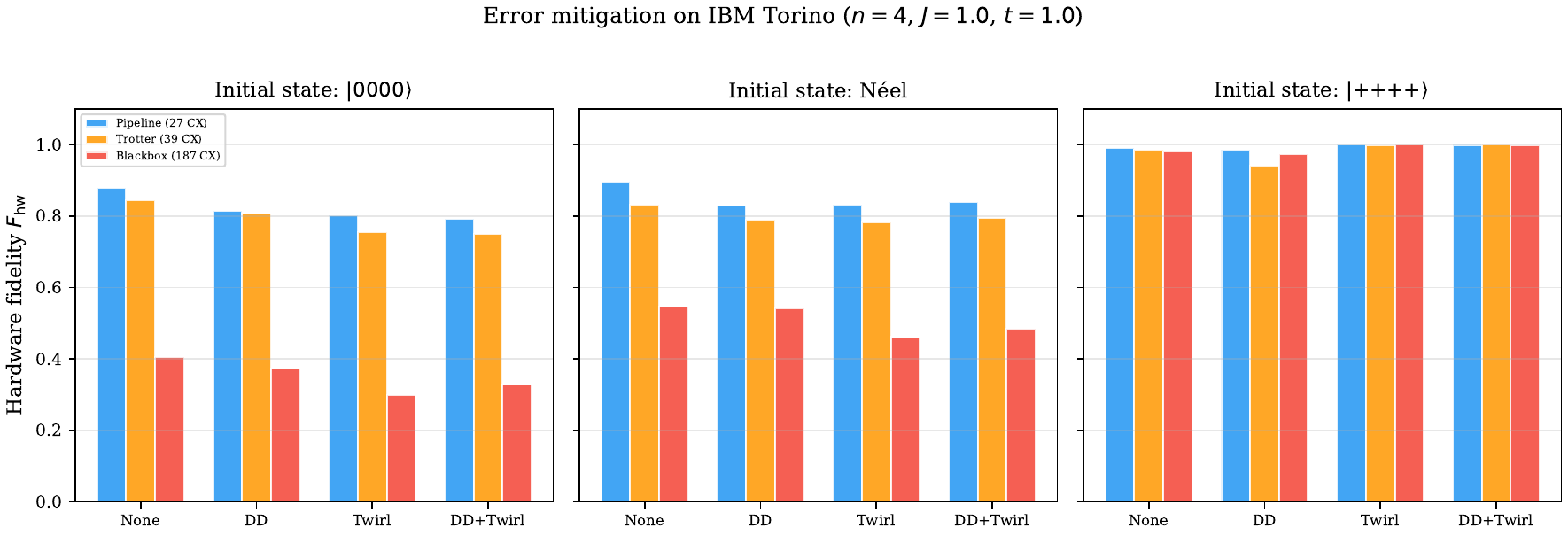}
\caption{Error mitigation comparison on IBM Torino ($n = 4$, $J = 1.0$). For computational basis states, mitigation provides no consistent improvement. For the $|++++\rangle$ state, twirling modestly improves all methods but does not change the relative ordering. The depth gap between pipeline (27 CX) and blackbox (187 CX) circuits cannot be closed by mitigation.}
\label{fig:mitigation}
\end{figure*}

We tested three error mitigation strategies on IBM Torino ($n = 4$, $J = 1.0$, $t = 1.0$) in a dedicated session separate from the baseline measurements in Table~\ref{tab:hw}; the no-mitigation baselines differ by $\sim$0.01--0.05 due to calibration drift between sessions (consistent with the session-to-session variability noted in Section~\ref{subsec:hardware}).
Bootstrap uncertainty estimates for this session are comparable to the main session ($\sigma_{F_{\mathrm{hw}}} \approx 0.003$--$0.008$); the differences between mitigation strategies (0.01--0.06) exceed this statistical uncertainty.
We evaluated dynamical decoupling (DD), Pauli twirling, and their combination (DD$+$twirl)~\cite{cai2023quantum}.
The results (Figure~\ref{fig:mitigation}) reveal an important finding: error mitigation techniques do not close the depth gap between pipeline and blackbox circuits.

For computational basis states ($|0000\rangle$, N\'{e}el), mitigation techniques provide negligible benefit and sometimes degrade performance.
DD decreases pipeline fidelity on $|0000\rangle$ from $F = 0.879$ to $F = 0.815$ because DD inserts additional identity-preserving gate sequences (we used the XX sequence, Qiskit's default for \texttt{PadDynamicalDecoupling}) between operations, increasing total circuit depth without reducing depolarizing noise, the dominant error source at $\sim 5 \times 10^{-3}$ per CX on current IBM hardware~\cite{viola1999dynamical}.
For short circuits such as our 27-CX pipeline, the marginal noise from DD's added gates outweighs any benefit from coherent error suppression.
Alternative DD sequences such as XpXm may offer marginally different performance, but since the dominant error source for our short circuits is depolarizing noise rather than coherent $Z$-errors, we do not expect a qualitative change in the conclusions.
Twirling similarly converts coherent errors to stochastic ones but cannot reduce the depolarizing floor, and thus degrades performance on computational basis states where total depth dominates.

For the superposition state $|++++\rangle$, twirling provides a modest improvement: pipeline $F$ increases from $0.989$ to $0.999$, and blackbox $F$ from $0.980$ to $0.999$.
However, this improvement is state-specific and does not generalize to computational basis inputs.

The key finding is that for our pipeline's already-short circuits (27 CX), error mitigation has limited room for improvement, while for the blackbox's deep circuits (187 CX), mitigation cannot compensate for the accumulated noise.
\emph{Circuit depth reduction is more effective than post-hoc error mitigation}---a practical conclusion consistent with the NISQ paradigm.

\section{Discussion}
\label{sec:discussion}

The efficacy of the proposed approach can be understood as a consequence of aligning the compilation procedure with the geometric structure of Hamiltonian evolution. Generic synthesis treats the target unitary as an arbitrary element of $\mathrm{SU}(2^n)$, leading to constructions whose cost grows rapidly with system size. In contrast, the present method restricts the search to a submanifold generated by local Hamiltonian terms, where meaningful approximations can be achieved with a number of parameters and gates that scale linearly with system size. This structural alignment explains both the efficiency of the greedy block selection and the favorable optimization landscape of the variational stage.

\subsection{Why physics-informed blocks work}
\label{subsec:why_blocks}

The effectiveness of Trotter--Suzuki blocks for compiling Hamiltonian unitaries has a geometric explanation.
The unitary $U = e^{-iHt}$ lies on a geodesic in $\mathrm{SU}(2^n)$ determined by the generator $H$.
Individual gates displace the compiled unitary along low-dimensional subspaces of $\mathrm{SU}(2^n)$, producing fidelity improvements of order $d^{-1} = 2^{-n}$, exponentially vanishing with system size (a single two-qubit gate acts non-trivially on a $4$-dimensional subspace of the $d$-dimensional Hilbert space, so its contribution to $|\mathrm{Tr}(U^{\dagger}V)|/d$ is concentrated in an $O(4/d)$ fraction of the terms, yielding fidelity changes of order $O(d^{-1})$).
In contrast, each Trotter block advances along the \emph{same geodesic} as the target (up to Trotter error), producing fidelity improvements of $O(0.1\text{--}0.3)$ per step.

This geometric alignment explains the consistent preference for S$_2$ blocks across all tested regimes (varying $n$, $J$, $t$, and Hamiltonian type).
Local Strang blocks apply only 3 of the Hamiltonian's $3(n{-}1)$ coupling terms per block, advancing along a low-dimensional submanifold and yielding fidelity improvements an order of magnitude smaller than S$_2$ blocks that coordinate all terms simultaneously.
S$_4$ blocks achieve higher accuracy per step but at $4\times$ the CX cost per block (Section~\ref{subsec:s2_vs_s4}), making them Pareto-dominated by multiple S$_2$ steps in the CX-constrained NISQ regime.

The effective candidate space thus collapses from the full library to a single block type, indicating that for Hamiltonian compilation, global Trotter steps dominate over local operations.
This is a direct consequence of the geometric structure: gate-level compilation requires $O(n^2)$ elementary gates per fidelity improvement step (each contributing $O(2^{-n})$ fidelity gain), whereas block-level compilation requires $O(1)$ blocks per step (each contributing $O(0.1$--$0.3)$ fidelity gain).
The block-based formulation thus reduces the effective search space to $O(1)$, making greedy selection sufficient without sophisticated optimization strategies.

This observation highlights that the advantage of physics-informed compilation is not merely quantitative but structural, i.e., by operating in a space aligned with the generator of the dynamics, the synthesis process avoids the exponentially inefficient exploration inherent to generic unitary decomposition.
\subsection{Pipeline components and routing advantage}
\label{subsec:roles}

Our ablation study (Table~\ref{tab:ablation}) reveals that the three pipeline components contribute differently across coupling regimes.

\textbf{Native circuit placement} provides the consistent, regime-independent advantage.
By decomposing Pauli rotations directly onto the hardware coupling map, it eliminates the SWAP routing overhead that inflates blackbox circuits by $3$--$136\times$ (Table~\ref{tab:scaling}).
Generic transpilation of a \texttt{UnitaryGate} produces an all-to-all gate decomposition requiring SWAP insertion (3 CX per SWAP via SABRE routing~\cite{li2019tackling}); our native circuits act on physically adjacent qubits by construction, requiring no SWAPs or qubit permutations.
This routing advantage grows with system size: at $n = 5$, blackbox transpilation requires 863 CX (versus 93 for our pipeline), much of it from SWAP routing.

This hardware alignment is particularly important in the NISQ regime, where routing overhead can dominate circuit depth. By ensuring that all entangling operations respect the device connectivity, the compiled circuits preserve the locality of the Hamiltonian while avoiding the accumulation of additional noise from SWAP operations.

\textbf{Variational refinement} is the critical component at strong coupling ($J \geq 1.0$).
In this regime, the Trotter error per step is large enough that neither greedy block selection nor fixed Trotterization with a reasonable number of steps can reach $F > 0.99$ without excessive CX cost.
Variational optimization of the Pauli rotation angles closes the gap: at $J = 1.0$, $n = 4$, it raises fidelity from $F = 0.55$ (greedy alone) to $F = 0.996$ with only 27 CX, compared to 63 CX for fixed $m = 5$ Trotterization.

From this perspective, the role of the variational stage is not to explore the full unitary space, but to locally refine a physically informed approximation. The Trotter-based initialization ensures that optimization begins in a region where gradients remain informative, enabling efficient convergence with a small number of parameters.

\textbf{Automated block selection} (the greedy oracle) provides both automation and, with the expanded candidate set (Appendix~\ref{app:dt_granularity}), CX reduction.
At weak coupling ($J \leq 0.5$), the oracle selects $m = 3$ blocks with $\Delta t = t/3$ for $t = 1.0$, matching the optimal Trotter discretization without user intervention.
The oracle's primary value is automation: it adapts $m$ and $\Delta t$ to the specific target without requiring the user to estimate the appropriate number of Trotter steps.

This further reinforces that the oracle acts primarily as an adaptive discretization mechanism rather than a general-purpose optimizer. By selecting from a set of physically meaningful blocks, it avoids the exponentially large search space associated with gate-level synthesis.
We note that Qiskit's own \texttt{PauliEvolutionGate} with \texttt{SuzukiTrotter} also benefits from structure-aware transpilation, and indeed our ``Qiskit Suzuki-2'' baseline uses the same physics as our native circuits.
The additional value of our pipeline lies in automated step selection and variational refinement at strong coupling.

\subsection{The universality--performance tradeoff}
\label{subsec:tradeoff}

Our results illustrate a fundamental tradeoff in quantum compilation.
Generic transpilers are designed for \emph{universality}: they decompose any unitary without knowledge of its origin.
This universality comes at a cost, the decomposition cannot exploit structural regularities, leading to circuits that are deeper than necessary for structured targets. Our pipeline sacrifices universality (it requires knowledge of $H$) in exchange for dramatically more efficient circuits.

This tradeoff is not a deficiency of either approach but a consequence of information theory: compilation with side information (the Hamiltonian) can always outperform compilation without it, just as lossy compression with a domain-specific model outperforms generic compression.
Importantly, the $14\times$ CX reduction over blackbox transpilation at $n = 4$ reflects two distinct effects: (i) structure-aware decomposition (shared by our method and Qiskit Suzuki-2, both using 15 CX) versus generic unitary synthesis ($\sim$100 CX for the raw decomposition), and (ii) elimination of SWAP routing overhead ($\sim$100 additional CX from SABRE).
Both effects are consequences of exploiting Hamiltonian structure: the first at the algebraic level (Trotter decomposition), the second at the topological level (native placement on the coupling map).
The practical implication is that quantum software stacks should expose domain-specific compilation pathways alongside generic transpilation.
For the increasingly important application domain of Hamiltonian simulation, encompassing quantum chemistry~\cite{peruzzo2014variational}, condensed matter physics~\cite{kim2023evidence}, and materials science~\cite{clinton2024towards}, physics-informed compilation provides a significant advantage over generic tools for local Hamiltonians on linear-connectivity hardware.

These results reveal a clear inversion of the conventional objective of compilation: maximizing simulation fidelity does not necessarily maximize hardware performance. Instead, reducing circuit depth—particularly the number of entangling gates—dominates the achievable fidelity on NISQ devices. In this regime, approximate circuits that respect Hamiltonian structure can outperform exact decompositions that ignore it.

We note that our CX counts depend on Qiskit's transpiler passes at \texttt{optimization\_level}$\geq 2$ for two-qubit block consolidation via KAK decomposition.
Alternative compilers, such as Quantinuum's \texttt{tket}~\cite{sivarajah2020tket}, which employs ZX-calculus-based rewriting and peephole optimizations, could potentially achieve comparable or greater consolidation.
We also note recent Pauli-network synthesis methods---Paulihedral~\cite{li2022paulihedral} and Rustiq~\cite{debrugiere2024rustiq}---that exploit Hamiltonian structure for circuit synthesis at the Pauli-rotation level.
These methods address a complementary optimization target (minimizing CX count within a fixed set of Pauli rotations) and could potentially be combined with our pipeline's adaptive block selection and variational refinement stages; a systematic comparison is left for future work.
Our pipeline is transpiler-agnostic at the circuit construction level: the native Pauli rotation sequences can be fed to any compiler that supports CX resynthesis of consecutive two-qubit blocks.
The 34$\to$15 CX reduction is a property of the palindromic block structure (which creates contiguous same-edge operations), not of a specific transpiler implementation, and we expect similar consolidation from any compiler with two-qubit block detection.

\subsection{Suzuki-2 versus Suzuki-4}
\label{subsec:s2_vs_s4}

A natural question is whether higher-order Trotter formulas would improve the pipeline.
Our experiments show that S$_4$ achieves marginally higher fidelity per step ($F = 0.99999$ at $m = 1$) but at $4\times$ the CX cost (63 versus 15 CX per block for $n = 4$ Heisenberg).
The S$_2$ formula with $m = 5$ steps (63 CX, $F = 0.99999$) matches S$_4$ with $m = 1$ in both CX count and fidelity.

For NISQ applications, where each CX gate incurs $\sim$0.5--1\% error, the lower-order formula with more steps is preferable because the CX gates are distributed across a shallower circuit with more intermediate error cancellation opportunities.
S$_4$ may become advantageous in the fault-tolerant regime where gate errors are negligible and Trotter accuracy is the limiting factor.

This comparison illustrates a broader principle that in the presence of noise, the optimal compilation strategy balances approximation error against implementation cost, favoring constructions that minimize entangling-gate count rather than those that minimize formal Trotter error.
\subsection{Limitations and outlook}
\label{subsec:limitations}

The present framework is designed for Hamiltonians whose structure can be directly leveraged during compilation, which defines both its strengths and its current scope of applicability.

\textbf{Requirement for Hamiltonian knowledge.}
The pipeline requires knowledge of $H$, making it inapplicable to arbitrary unitaries.
This is a fundamental restriction: the CX advantage derives from exploiting the structure of $H$, and no free lunch theorem guarantees that no method can simultaneously be universal and structure-optimal.
However, for the primary application domain---quantum simulation of known physical systems, the Hamiltonian is always available by definition.

\textbf{Exponential classical cost.}
The variational refinement stage relies on Pauli decomposition and full $2^n \times 2^n$ unitary computation, both scaling as $O(4^n)$.
This restricts practical application to $n \lesssim 8$ qubits; we emphasize that the present work is a proof-of-concept demonstrating the viability of structure-aware compilation for $n \leq 8$, not a scalable production tool.
Table~\ref{tab:classical_cost} in Appendix~\ref{app:classical_cost} summarizes the classical cost; L-BFGS optimization dominates at $n \leq 6$, while the $O(4^n)$ Pauli decomposition dominates at $n \geq 7$.
Scaling beyond $n = 8$ would require replacing the full unitary computation with scalable alternatives: tensor network-based state representations~\cite{vidal2003efficient, orus2019tensor}, parameter-shift rules~\cite{mitarai2018quantum} or adjoint differentiation~\cite{jones2020efficient} for circuit-level gradient computation, and block-diagonal approximations to reduce the Pauli decomposition cost.

\textbf{Topology.}
The pipeline has been validated on one-dimensional nearest-neighbor Hamiltonians, where hardware connectivity can be directly exploited. This regime captures a large class of experimentally relevant systems and maximizes the benefit of eliminating routing overhead. Extending the approach to more general connectivity graphs is a natural direction for future work.

\textbf{Fixed ansatz structure.}
The variational layer structure (Rz + XX/YY/ZZ per edge) is tailored to isotropic Heisenberg-type Hamiltonians.
For Hamiltonians with different symmetries or long-range interactions, the ansatz should be adapted accordingly.
Symmetry-adapted blocks~\cite{wiersema2020exploring} that respect conservation laws (e.g., total magnetization) could both reduce the parameter count and improve convergence.

\textbf{Outlook.}
Several extensions could broaden the pipeline's applicability: continuous time-step optimization within a single block (beyond the discrete $t/m$ grid), automated ansatz construction based on Hamiltonian symmetries, integration with randomized compilation techniques~\cite{campbell2019random} for improved Trotter accuracy without increased depth, extension to higher-order product formulas optimized for specific Hamiltonian structures~\cite{childs2019nearly}, and packaging the pipeline as a custom Qiskit transpiler pass for \texttt{PauliEvolutionGate} targets to make structure-aware compilation accessible within standard workflows.
More broadly, the present results suggest that compilation strategies for near-term quantum devices should be guided by physical structure and hardware constraints, rather than by exact unitary synthesis alone.
Recent large-scale Hamiltonian simulation experiments~\cite{kim2023evidence} underscore the practical importance of efficient Hamiltonian simulation circuits, and our pipeline provides a concrete path toward realizing this utility on near-term hardware.


\section{Conclusion}
\label{sec:conclusion}

We have introduced a structure-aware framework for compiling Hamiltonian dynamics into hardware-native quantum circuits. By combining native placement, adaptive product-formula synthesis, and Trotter-initialized variational refinement, the method produces shallow circuits that maintain high fidelity while respecting hardware constraints.

Our results demonstrate that, in the NISQ regime, approximate structure-preserving compilation can outperform exact unitary synthesis. This inversion arises because the accumulation of noise in deep circuits outweighs the benefits of exact decomposition. By reducing entangling-gate count and aligning the circuit structure with the underlying Hamiltonian, the proposed approach enables more reliable execution of quantum simulations on current devices.

The framework provides a practical pathway for near-term quantum simulation without requiring pulse-level control, and suggests a broader shift in compilation strategy toward physics-informed, hardware-aware methods.

\section*{Acknowledgments}

FSL thanks the S\~ao Paulo Research Foundation (FAPESP), Brasil. Process Number \mbox{2026/04387-7} for financial support and Prof. Felipe Fernandes Fanchini for lending computing resources used for the classical optimization and simulation experiments.
MCO acknowledges partial financial support from the National Institute of Science and Technology for Applied Quantum Computing through CNPq process No. 408884/2024-0 and by FAPESP, through the Center for Research and Innovation on Smart and Quantum Materials (CRISQuaM) process No. 2013/07276-1.
The authors acknowledge use of the IBM Quantum platform for hardware validation.
The views expressed are those of the author and do not reflect the official policy or position of IBM or the IBM Quantum team.

\section*{Conflict of Interest}

The author declares no conflict of interest.

\section*{Data Availability Statement}

The code and data that support the findings of this study are available at \href{https://github.com/fsluiz/efficient-hamiltonian-synthesis}{github}
\appendix

\section{Reproducible gate counts}
\label{app:gate_counts}

This appendix provides full evidence for the CX gate counts reported in the main text.
All results are generated by a self-contained script (\texttt{verify\_cx\_counts.py}) using Qiskit 2.3.0 with the following fixed settings:
\texttt{basis\_gates=['cx','rz','sx','x']},
\texttt{coupling\_map} = linear chain $[0\text{-}1, 1\text{-}2, 2\text{-}3]$,
\texttt{initial\_layout=[0,1,2,3]} (identity, no SABRE routing),
and \texttt{seed\_transpiler=42}.
Results are verified to be identical across 10 different transpiler seeds (0--9).

\subsection{\texorpdfstring{Raw versus transpiled circuit for a single S\textsubscript{2} block}{Raw versus transpiled circuit for a single S2 block}}

The Suzuki-2 palindromic iteration for the Heisenberg model on $n = 4$ qubits (9 coupling terms, 0 field terms) generates 17 rotation steps, each requiring 2 CX gates (Eq.~\ref{eq:zz_decomp}), for a total of \textbf{34 raw CX}.
The palindromic ordering groups rotations by edge:
\begin{itemize}
\item Edge $(0,1)$ (outermost): 10 raw CX.
\item Edge $(1,2)$ (interior): 12 raw CX.
\item Edge $(2,3)$ (innermost, contains palindrome center): 12 raw CX.
\end{itemize}

Table~\ref{tab:transpile_levels} shows the CX count after transpilation at each optimization level.
At \texttt{opt\_level}$\geq 2$, Qiskit's \texttt{Collect2qBlocks} and \texttt{ConsolidateBlocks} passes detect consecutive two-qubit operations on the same qubit pair and resynthesize each block as an optimal two-qubit unitary via KAK decomposition~\cite{khaneja2001cartan}, requiring at most 3 CX per block.
The large reduction in total gate count (243 at \texttt{opt\_level}$= 0$ to 67 at \texttt{opt\_level}$= 2$) reflects both CX consolidation ($34 \to 15$, a 19-gate reduction) and single-qubit gate merging ($209 \to 52$ single-qubit gates, as consecutive Rz and basis-change gates on the same qubit are absorbed into the resynthesized two-qubit blocks).

\begin{table}[ht]
\centering
\caption{CX gate count per optimization level for a single S$_2$ block ($n = 4$, Heisenberg, $J = 0.5$, $\Delta t = 0.2$, Qiskit 2.3.0). Results are seed-independent (identical for seeds 0--9).}
\label{tab:transpile_levels}
\begin{tabular}{ccccccc}
\toprule
\texttt{opt\_level} & Total CX & Depth & Total gates & \multicolumn{3}{c}{CX per edge} \\
\cmidrule(lr){5-7}
& & & & $(0,1)$ & $(1,2)$ & $(2,3)$ \\
\midrule
0 & 34 & 147 & 243 & 10 & 12 & 12 \\
1 & 34 & 96 & 141 & 10 & 12 & 12 \\
2 & 15 & 42 & 67 & 3 & 6 & 6 \\
3 & 15 & 42 & 67 & 3 & 6 & 6 \\
\bottomrule
\end{tabular}
\end{table}

The per-edge counts at \texttt{opt\_level}$= 3$ have a structural explanation.
Edge $(0,1)$ is the \emph{outermost} edge in the palindromic ordering: its forward rotations (steps 0--2) and backward rotations (steps 14--16) form a single contiguous block of 10 CX with no intervening operations on other edges.
The transpiler consolidates this into one optimal 2-qubit unitary $\to$ 3 CX.
Edges $(1,2)$ and $(2,3)$ are \emph{interior}: their forward and backward groups are separated by the palindrome's middle section, creating two separate 2-qubit blocks per edge $\to$ $2 \times 3 = 6$ CX each.
Total: $3 + 6 + 6 = 15$ CX.

\subsection{Multi-block scaling formula}

Table~\ref{tab:multi_block} verifies the formula $\mathrm{CX}(m) = 15 + 12(m-1)$ for $m = 1$ to $10$ blocks.
At block boundaries, the last rotation of block $k$ and the first rotation of block $k{+}1$ act on the same qubit pair (edge $(0,1)$), enabling an additional consolidation.
Each boundary saves $15 - 12 = 3$ CX relative to independent blocks.

\begin{table*}[ht]
\centering
\caption{CX gate count for $m$ concatenated S$_2$ blocks ($n = 4$, Heisenberg, $t = 1.0$, Qiskit 2.3.0, \texttt{opt\_level=3}). The formula $\mathrm{CX}(m) = 15 + 12(m-1)$ is exact for all tested values.}
\label{tab:multi_block}
\begin{tabular}{cccccc}
\toprule
$m$ & CX (raw) & CX (transpiled) & CX (formula) & Match & Process fidelity $F$ \\
\midrule
1 & 34 & 15 & 15 & \checkmark & 0.99997 \\
2 & 68 & 27 & 27 & \checkmark & 0.99999 \\
3 & 102 & 39 & 39 & \checkmark & 0.99999 \\
4 & 136 & 51 & 51 & \checkmark & 0.99999 \\
5 & 170 & 63 & 63 & \checkmark & 0.99999 \\
6 & 204 & 75 & 75 & \checkmark & 0.99999 \\
7 & 238 & 87 & 87 & \checkmark & 0.99999 \\
8 & 272 & 99 & 99 & \checkmark & 0.99999 \\
9 & 306 & 111 & 111 & \checkmark & 1.00000 \\
10 & 340 & 123 & 123 & \checkmark & 1.00000 \\
\bottomrule
\end{tabular}
\end{table*}

\subsection{Variational layer gate cancellations}

Each variational layer (Rz on $n$ qubits + XX, YY, ZZ on $n-1$ edges) contains $6(n-1) = 18$ raw CX for $n = 4$.
After transpilation (\texttt{opt\_level=3}), the three Pauli rotations per edge (XX, YY, ZZ) are consolidated into a single optimal 2-qubit unitary (3 CX per edge), giving $3(n-1) = 9$ CX per layer.
For $L$ layers, CX$= 9L$ (verified for $L = 1$--$5$: 9, 18, 27, 36, 45 CX respectively).

\subsection{Cross-Hamiltonian verification}

Table~\ref{tab:cross_ham} shows that the transpilation-based reduction is consistent across Hamiltonian types.
For Ising (ZZ-only couplings), no cancellation occurs because each edge has only one type of rotation; for XY (XX+YY per edge), the two rotations per edge consolidate into a single 2-qubit unitary. The field count on Ising refers to the X Pauli, one for each qubit.

\begin{table*}[ht]
\centering
\caption{CX count for a single S$_2$ block across Hamiltonian types ($n = 4$, $J = 0.5$, $\Delta t = 0.2$, \texttt{opt\_level=3}).}
\label{tab:cross_ham}
\begin{tabular}{lccccc}
\toprule
Hamiltonian & Couplings & Fields & CX (raw) & CX (transpiled) & Reduction \\
\midrule
Heisenberg & 9 (XX+YY+ZZ) & 0 & 34 & 15 & 2.3$\times$ \\
XY & 6 (XX+YY) & 0 & 22 & 10 & 2.2$\times$ \\
Ising & 3 (ZZ) & 4 & 10 & 10 & 1.0$\times$ \\
\bottomrule
\end{tabular}
\end{table*}

\subsection{Suzuki-4 gate count verification}

For completeness, a single S$_4$ block ($n = 4$, Heisenberg) contains $5 \times 34 = 170$ raw CX (five S$_2$ sub-blocks per Eq.~\ref{eq:suzuki4}).
After transpilation at \texttt{opt\_level=3}, this reduces to 63 CX (depth 178, 273 total gates)---a $2.7\times$ reduction.
This matches the S$_2$ formula at $m = 5$ blocks (both yield 63 transpiled CX), confirming that S$_4$ with $m = 1$ and S$_2$ with $m = 5$ have identical CX cost, as noted in Section~\ref{subsec:s2_vs_s4}.
At \texttt{opt\_level} $\leq 1$, all 170 raw CX are retained.

\section{Fidelity metrics}
\label{app:fidelity_metrics}

This work uses two distinct fidelity metrics that are \emph{not} numerically comparable:

\begin{enumerate}
\item \textbf{Process fidelity} $F(U, V) = |\mathrm{Tr}(U^{\dagger}V)|^2/d^2$ (Eq.~\ref{eq:fidelity}): compares two $d \times d$ unitary matrices.
This metric is used for all simulation-level comparisons (Tables~\ref{tab:comparison}--\ref{tab:noise}, Figures~\ref{fig:sweep}--\ref{fig:noise}).
It relates to the average gate fidelity via $F_{\mathrm{avg}} = (dF + 1)/(d+1)$; for $d = 16$ ($n = 4$), $F = 0.997$ corresponds to $F_{\mathrm{avg}} = 0.9972$.

\item \textbf{Hellinger fidelity} $F_{\mathrm{hw}} = (\sum_x \sqrt{p_x q_x})^2$: compares two classical probability distributions over measurement outcomes.
This metric is used for all hardware results (Table~\ref{tab:hw}, Figures~\ref{fig:hardware}--\ref{fig:mitigation}).
It is sensitive to state-preparation-and-measurement (SPAM) errors that do not affect process fidelity.
A value $F_{\mathrm{hw}} = 0.99$ does \emph{not} imply $F \approx 0.99$, as the two metrics probe fundamentally different quantities (distribution overlap versus unitary proximity).
\end{enumerate}

\section{Greedy oracle and time-step granularity}
\label{app:dt_granularity}

The greedy oracle's performance depends on the granularity of the candidate time-step set.
With the base library $\Delta t \in \{0.05, 0.1, 0.2\}$, the maximum step is $\Delta t_{\max} = 0.2$, which forces at least $\lceil t / 0.2 \rceil = 5$ blocks for $t = 1.0$.
This is a limitation of the candidate set, not of the method: the oracle greedily selects $\Delta t = 0.2$ at each step (producing $m = 5$, 63 CX for $n = 4$ Heisenberg), even though a single Trotter step with $\Delta t = t/3 = 0.333$ would achieve comparable fidelity with fewer blocks.

Expanding the candidate set to include standard Trotter discretizations $\Delta t = t/m$ for $m \geq 3$ resolves this limitation.
Table~\ref{tab:dt_granularity} shows the effect: with the expanded set, the greedy oracle discovers $m = 3$ ($\Delta t = 0.333$, 39 CX), a 38\% reduction from the default 63 CX, with negligible fidelity loss ($F = 0.9999$ for both).
At $m = 2$ ($\Delta t = 0.5$, 27 CX), fidelity remains above 0.999.
Including $\Delta t = t$ ($m = 1$) triggers a greedy myopia failure: the oracle selects the single large step (largest immediate fidelity gain) but then cannot improve further, stalling at $F = 0.991$.

\begin{table}[ht]
\centering
\caption{Effect of candidate $\Delta t$ granularity on greedy oracle performance ($n = 4$, Heisenberg, $J = 0.5$, $t = 1.0$). The expanded set $\{t/m : m \geq 3\}$ reduces the block count from 5 to 3 without meaningful fidelity loss.}
\label{tab:dt_granularity}
\begin{tabular}{lcccc}
\toprule
Candidate set & $\Delta t_{\max}$ & Blocks & CX & $F$ \\
\midrule
$\{0.05, 0.1, 0.2\}$ (base) & 0.200 & 5 & 63 & 0.99998 \\
$+\{t/m : m \geq 3\}$ & 0.333 & 3 & 39 & 0.99987 \\
$+\{t/m : m \geq 2\}$ & 0.500 & 2 & 27 & 0.99935 \\
$+\{t/m : m \geq 1\}$ & 1.000 & 2\textsuperscript{a} & 27 & 0.99133 \\
\bottomrule
\end{tabular}

\smallskip
\noindent\textsuperscript{a}Greedy myopia: the oracle selects $\Delta t = 1.0$ (reaching $F = 0.991$, below the $0.999$ target), then attempts a second block ($\Delta t = 0.05$) to close the gap, but the incremental improvement is below the $10^{-6}$ threshold, triggering termination. The oracle does not enforce $\sum_k \Delta t_k = t$ as a constraint---it treats each block as an ansatz element whose angles are determined by the Trotter formula.
\end{table}

Across 40 parameter configurations (Section~\ref{subsec:sweep} grid plus qubit scaling), expanding the candidate set to include $\{t/m : m \geq 3\}$ improved 18 cases (reducing CX by 16--84), left 21 unchanged, and produced one regression (Heisenberg $J = 1.0$, $t = 2.0$: an extreme case where $F_{\mathrm{base}} = 0.11$).
For the qubit scaling series ($n = 3$--$6$, $J = 0.5$), the expanded set consistently reduces CX by $\sim$38\% (e.g., from 123 to 75 CX at $n = 6$) with $F > 0.9998$.
At strong coupling ($J = 1.0$), the expanded set has no effect because the greedy oracle selects zero blocks regardless of the candidate set, and variational refinement handles the compilation.

Hardware validation on IBM Torino confirms this improvement.
A dedicated run with the expanded $\Delta t$ library yielded $F_{\mathrm{hw}} = 0.77$ for the $m = 3$ pipeline (39 CX) on the $|0\rangle$ state at $n = 4$, $J = 0.5$, compared to $F_{\mathrm{hw}} = 0.60$ for the $m = 5$ pipeline (63 CX) from the original session---a +17 percentage point improvement, attributable to the 38\% reduction in two-qubit gate count.
We recommend including $\{t/m\}$ values in the candidate set as the default configuration.

\section{Barren plateau data}
\label{app:barren_data}

Table~\ref{tab:barren} provides the full gradient landscape data underlying Figure~\ref{fig:barren}.

\begin{table}[ht]
\centering
\caption{Gradient landscape analysis for the Trotter-structured variational ansatz ($L = 3$ layers, Heisenberg $J = 1.0$, 100 samples $\times$ 10 parameters). Under random initialization, gradient variance decays exponentially ($\sim 0.067^n$, barren plateau). Under Trotter initialization, gradients remain 1--3 orders of magnitude larger, enabling robust optimization up to $n = 8$.}
\label{tab:barren}
\begin{tabular}{cccccc}
\toprule
$n$ & Params & $\mathrm{Var}[\frac{\partial C}{\partial\theta}]_{\mathrm{rand}}$ & $\mathrm{Var}[\frac{\partial C}{\partial\theta}]_{\mathrm{Trotter}}$ & $\langle|\frac{\partial C}{\partial\theta}|\rangle_{\mathrm{Trotter}}$ & Ratio\textsuperscript{c} \\
\midrule
3 & 27 & $1.7 \times 10^{-3}$ & $2.3 \times 10^{-2}$ & $1.1 \times 10^{-1}$ & $4\times$ \\
4 & 39 & $9.3 \times 10^{-5}$ & $9.8 \times 10^{-3}$ & $7.6 \times 10^{-2}$ & $12\times$ \\
5 & 51 & $5.5 \times 10^{-6}$ & $2.2 \times 10^{-3}$ & $3.8 \times 10^{-2}$ & $24\times$ \\
6 & 63 & $4.0 \times 10^{-7}$ & $1.7 \times 10^{-3}$ & $3.6 \times 10^{-2}$ & $92\times$ \\
7 & 75 & $2.5 \times 10^{-8}$ & $1.2 \times 10^{-3}$ & $3.0 \times 10^{-2}$ & $305\times$ \\
8 & 87 & $2.3 \times 10^{-9}$ & $1.1 \times 10^{-3}$ & $2.8 \times 10^{-2}$ & $1{,}011\times$ \\
\bottomrule
\end{tabular}

\smallskip
\noindent\textsuperscript{c}Ratio $= \mathrm{Var}_{\mathrm{Trotter}} / \mathrm{Var}_{\mathrm{rand}}$, measuring the gradient variance advantage of Trotter initialization over random initialization.
\end{table}

\section{Classical computational cost}
\label{app:classical_cost}
Table \ref{tab:classical_cost} presents the classical cost.
\begin{table}[ht]
\centering
\caption{Classical cost of the full pipeline (greedy selection + variational refinement + transpilation) per instance, single-threaded (no GPU acceleration). L-BFGS iterations and function evaluations are for the variational stage at $J = 1.0$; $L$ denotes the number of variational layers required for $F > 0.99$.
Matrix exponentials use pre-computed eigendecompositions for $O(d^2)$ per evaluation.}
\label{tab:classical_cost}
\begin{tabular}{cccccc}
\toprule
$n$ & $L$ & Wall-clock time & L-BFGS iters & Func. evals & Bottleneck \\
\midrule
4 & 3 & $\sim$5\,s & $\sim$200 & $\sim$9\,200 & L-BFGS \\
5 & 5 & $\sim$35\,s & $\sim$250 & $\sim$15\,000 & L-BFGS \\
6 & 5 & $\sim$2.5\,min & $\sim$250 & $\sim$15\,000 & L-BFGS \\
7 & --- & $\sim$2\,min\textsuperscript{d} & ---\textsuperscript{d} & ---\textsuperscript{d} & Pauli decomp. \\
8 & --- & $\sim$9\,min\textsuperscript{d} & ---\textsuperscript{d} & ---\textsuperscript{d} & Pauli decomp. \\
\bottomrule
\end{tabular}

\smallskip
\noindent\textsuperscript{d}Variational refinement was not run at $n \geq 7$; the wall-clock times reflect greedy block selection and Pauli decomposition only.
At $n = 6$, the success rate is 3/5 seeds; at $n \geq 7$, the gradient variance ($\sim 0.54^n$) and parameter count ($> 125$) make convergence unreliable.
For comparison, Qiskit's blackbox transpilation (\texttt{UnitaryGate}, \texttt{optimization\_level=3}) takes $< 1$\,s at $n = 4$--$6$; our pipeline is slower due to the L-BFGS optimization stage, but the resulting circuits are $14$--$136\times$ shorter, which is the relevant metric for NISQ hardware execution.
\end{table}

\section{FakeTorino noise model validation}
\label{app:faketorino}

Table~\ref{tab:faketorino} presents simulation results using the FakeTorino noise model (Qiskit Aer), which incorporates gate-dependent error rates, amplitude damping, phase damping, and crosstalk derived from IBM Torino calibration data.
These results bridge the gap between the idealized depolarizing model (Section~\ref{subsec:noise}) and real hardware (Section~\ref{subsec:hardware}).
The ordering of methods by $F_{\mathrm{hw}}$ is consistent with the real hardware results in Table~\ref{tab:hw}: shorter circuits consistently achieve higher fidelity under realistic noise.

\begin{table}[ht]
\centering
\caption{FakeTorino noise model results ($N = 8{,}192$ shots). Scenario A: $n = 4$, $J = 0.5$. Scenario B: $n = 4$, $J = 1.0$. Scenario C: $n = 5$, $J = 0.5$. The method ordering matches real hardware (Table~\ref{tab:hw}).}
\label{tab:faketorino}
\begin{tabular}{llcccc}
\toprule
& Method & CX & $F_{\mathrm{sim}}$ & $F_{\mathrm{hw}}$ ($|0\rangle^{\otimes n}$) & $F_{\mathrm{hw}}$ (N\'{e}el) \\
\midrule
\multicolumn{6}{l}{\textit{Scenario A ($n = 4$, $J = 0.5$)}} \\
& Trotter $m{=}3$ & 39 & 0.9999 & 0.817 & 0.853 \\
& Pipeline (greedy, $m{=}5$) & 63 & 1.0000 & 0.756 & 0.804 \\
& Blackbox & 187 & 1.0000 & 0.469 & 0.616 \\
\midrule
\multicolumn{6}{l}{\textit{Scenario B ($n = 4$, $J = 1.0$)}} \\
& Pipeline (var$_3$) & 27 & 0.9955 & 0.856 & 0.871 \\
& Trotter $m{=}3$ & 39 & 0.9944 & 0.820 & 0.841 \\
& Trotter $m{=}5$ & 63 & 0.9993 & 0.756 & 0.780 \\
& Blackbox & 187 & 1.0000 & 0.456 & 0.603 \\
\midrule
\multicolumn{6}{l}{\textit{Scenario C ($n = 5$, $J = 0.5$)}} \\
& Pipeline (greedy, $m{=}5$) & 93 & 1.0000 & 0.676 & 0.734 \\
& Blackbox & 863 & 1.0000 & 0.056 & 0.275 \\
\bottomrule
\end{tabular}
\end{table}

\bibliography{references}

\end{document}